\documentclass[11pt,a4paper]{article}

\usepackage{epstopdf}% To incorporate .eps illustrations using PDFLaTeX, etc.
\usepackage{subfigure}% Support for small, `sub' figures and tables

\usepackage{enumitem} 
\usepackage{graphicx,hyperref,amsmath,natbib,bm,url}
\usepackage[utf8]{inputenc} % include umlaute

\usepackage{color}  
\definecolor{darkblue}{rgb}{0.15,0,0.37}
\definecolor{darkred}{rgb}{0.35,0,0.08}
\definecolor{mygrey}{rgb}{0.85,0.85,0.85} % make grey for the table
\usepackage{natbib}% Citation support using natbib.sty
\bibpunct[, ]{(}{)}{;}{a}{}{,}% Citation support using natbib.sty
\usepackage{tfrupee}
\usepackage{microtype,todonotes}
\usepackage[australian]{babel} % change date to to European format

\usepackage{caption}
\usepackage{longtable, booktabs, tabularx} % for nice tables
\usepackage{caption,fixltx2e}  % for nice tables
\usepackage[flushleft]{threeparttable}  % for nice tables

\usepackage{multicol}
\usepackage{etoolbox}
\usepackage{relsize}
\usepackage{blindtext}
\usepackage{geometry}
\usepackage[onehalfspacing]{setspace} % Line spacing

\usepackage[marginal]{footmisc} % footnotes not indented
\usepackage{pdflscape} % for landscape figures

\usepackage[capposition=top]{floatrow} % For notes below figure

\usepackage{color, colortbl}
\usepackage{multirow}
\usepackage{fancyhdr}
\usepackage{lastpage}
\usepackage{psfrag}

\usepackage{sepfootnotes}

\usepackage{rotating}
\usepackage{adjustbox}

\usepackage{hyperref}
\hypersetup{
	colorlinks=true,
	hypertexnames=false,
	hyperfootnotes=false,
	colorlinks,		
	linkcolor={red},
	citecolor={blue},
	urlcolor={black},
	pdfstartview={FitV},
	unicode,
	breaklinks=true
}

\makeatletter
\newcommand*{\indep}{%
	\mathbin{%
		\mathpalette{\@indep}{}%
	}%
}
\newcommand*{\nindep}{%
	\mathbin{%                   % The final symbol is a binary math operator
		\mathpalette{\@indep}{\not}% \mathpalette helps for the adaptation
	}%
}
\newcommand*{\@indep}[2]{%
	\sbox0{$#1\perp\m@th$}%        box 0 contains \perp symbol
	\sbox2{$#1=$}%                 box 2 for the height of =
	\sbox4{$#1\vcenter{}$}%        box 4 for the height of the math axis
	\rlap{\copy0}%                 first \perp
	\dimen@=\dimexpr\ht2-\ht4-.2pt\relax
	\kern\dimen@
	{#2}%
	\kern\dimen@
	\copy0 %                       second \perp
} 

\numberwithin{equation}{section}

\renewenvironment{abstract}
{\small
	\begin{center}
		\bfseries \abstractname\vspace{-.5em}\vspace{0pt}
	\end{center}
	\list{}{
		\setlength{\leftmargin}{.5cm}%
		\setlength{\rightmargin}{\leftmargin}%
	}%
	\item\relax}
{\endlist}

\newcommand{\pr}{\mathrm{Pr}}
\newcommand{\E}{\mathrm{E}}

\DeclareMathAlphabet{\mathbbold}{U}{bbold}{m}{n}

\renewcommand{\thetable}{\arabic{table}}
\include{commands}

\begin{document}
	
	%\vspace*{0.5cm}
	
	%\thispagestyle{empty} % page number not shown on first page

\begin{titlepage}
	\title{Productivity Implications of R\&D, Innovation, and Capital Accumulation for Incumbents and Entrants: Perspectives from a Catching-up Economy\thanks{This work was supported by the  European Union Horizon 2020 Research and Innovation Action under Grant Number 822781 and Estonian Research Council under Grant number PRG791. We owe thanks to Statistics Estonia for their indispensable help in supplying the data. The authors also acknowledge support for the compilation of the datasets used in the paper from the Estonian Research Infrastructures Roadmap project `Infotechnological Mobility Observatory (IMO).' We thank the participants of the $ 5^{\text{th}} $ RIE Conference and the $ 12^{\text{th}} $ Conference on Model-based Evidence on Innovation and Development (MEIDE) Maastricht, 12/2021, and the participants of the Baltic Economic Association Conference, Tartu, 06/2021. Thanks are due to Linas Tarasonis and Priit Vahter for providing valuable comments. The usual disclaimers apply.}}
	\author{Jaan Masso\thanks{Jaan Masso: University of Tartu,  jaan.masso@.ut.ee}, Amaresh K Tiwari \thanks{Amaresh K Tiwari: University of Tartu, amaresh.tiwari@ut.ee \&  amaresh.kr.tiwari@gmail.com} }
	\date{}
	\maketitle
	\begin{center}
		
	\end{center}
	\begin{abstract}
		 We study the productivity implications of R\&D, capital accumulation, and innovation output for entrants and incumbents in Estonia. First, in contrast to developed economies, a small percentage of firm engage in formal R\&D, but a much larger percentage innovate. Second, while we find no difference in the R\&D elasticity of productivity for the entrants and incumbents, the impact of innovation output --- many of which are a result of `doing, using and interacting' (DUI) mode of innovation --- is found to be higher for the entrants. Entrants who innovate are 21\% to 30\% more productive than entrants who do not; the corresponding figures for the incumbents are 10\% to 13\%. Third, despite the adverse sectoral composition typical of catching-up economies, Estonian incumbents, who are the primary carriers of `scientific and technologically-based innovative' (STI) activities, are comparable to their counterparts in developed economies in translating STI activities into productivity gains. Fourth, while embodied technological change through capital accumulation is found to be more effective in generating productivity growth than R\&D, the effectiveness is higher for firms engaging in R\&D. Finally, our results suggest that certain policy recommendations for spurring productivity growth in technologically advanced economies may not be applicable for catching-up economies. \\
		\vspace{0in}
		\vspace{0in}
	
		\bigskip
	\end{abstract}
	\setcounter{page}{0}
	\thispagestyle{empty}
\end{titlepage}

	\def\onepc{$^{\ast\ast\ast}$} \def\fivepc{$^{\ast\ast}$}
	\def\tenpc{$^{\ast}$}
	\def\legend{\multicolumn{3}{l}{\footnotesize{Significance levels
				:\hspace{1em} $\ast$ : 10\% \hspace{1em} $\ast\ast$ : 5\%
				\hspace{1em} $\ast\ast\ast$ : 1\% \normalsize}}}

	\defcitealias{doraszelski:2013}{DJ}
	\defcitealias{lubczyk:2020}{LP}
	\defcitealias{buehlmaier:2018}{BW}
	\defcitealias{olley:1996}{OP}
	\defcitealias{kinne:2018}{KA}
	\defcitealias{ackerberg:2015}{ACF} 
	\defcitealias{gandhi:2020}{GNR}
	\defcitealias{crepon:1998}{CDM}
	\defcitealias{lentz:2008}{LM}
	\defcitealias{levinsohn:2003}{LePe}

\section{Introduction}
	
The decline in productivity growth and job reallocation coupled with rising intra- and inter-industry dispersion in productivity during the last two decades in the major advanced economies have, as pointed out by \citet{akcigit:2021}, has coincided with decline in firm entry rate and young firms' share in economic activity. Since creative destruction --- the process by which newly innovated technologies replace older technologies --- plays a central role in many theories of growth, and since it is mainly attributed to innovation activities of fast growing, young innovative firms, the decline in productivity growth and economic activities of the young firms have prompted many recent studies on the innovative behaviour of the entrants and incumbents. 

While there are many such studies for the developed economies --- a partial list includes \citet{czarnitzki:2012}, \citet{coad:2016}, \citet{acemoglu:2018}, \citet{garcia:2019}, \citet{lubczyk:2020}, and \citet{akcigit:2021}  --- studies on innovative behaviour of entrants and incumbents in catching-up economies in Central and Eastern Europe (CEE) are lacking. Studying the innovative behaviour of these groups of firms in catching-up economies, which are at some distance from the technological frontier, is important not only because their National Innovation System (NIS) --- the web of linkages that direct flow of technology and information among actors such as  firms, universities and government research institutes --- is distinct from those of the developed economies \citep{radosevic:1998,castellacci:2013,cirillo:2019},\footnote{Notwithstanding the various theorizations of the NIS and the continuing work to develop a more robust concept of NIS \citep{cirillo:2019}, in various studies, such as \citet{castellacci:2013} and \citet{cirillo:2019}, catching-up economies of the Central and East Europe are categorized separately from the Western European countries and the US.} but also because policy prescriptions, as we show, for spurring growth in developed economies may not be applicable to the catching-up economies. Also, assessing the productivity implications of the firms' innovative behaviour can inform how the two groups of firms contribute to the aggregate productivity growth.             

%Even as work on NIS and the developed indicators to map knowledge flows has ``provided policymakers with an effective analytical toolkit, and has contributed to putting innovation policies centre stage on growth agendas," \citep{cirillo:2019}  argue that ``the concept remains elusive and difficult to articulate clearly, and even distilling an agreed definition from the literature is far from straightforward."

An outcome of Estonia's --- a representative CEE catching-up economy ---  distinctive NIS vis-\`{a}-vis developed economies is that, in our sample, about 30\% of the firm-years have positive R\&D expenses, but 74\% of the firm-years have innovated to introduce at least one new product in the market or a new process. This suggests that in Estonia more firms innovate through the  `doing, using and interacting' (DUI) mode of innovation rather than `scientific and technologically-based innovation' (STI) \citep[see][]{jensen:2007}. Moreover, as argued in \citet{peters1:2017}, while investments in R\&D increase the probability of realising product or process innovations, R\&D investment is neither necessary nor sufficient for firm innovation. Focusing on R\&D alone will, therefore, give a partial picture of the productivity implications of innovative activities \citep[see also][on the inadequacy of existing measures, that are overly R\&D oriented, for gauging technological change in middle-income economies]{radosevic:2019}. We, therefore, in contrast to some of the papers mentioned above, develop an empirical strategy to \textit{simultaneously} estimate (1) the productivity elasticity of own R\&D for the entrants and the incumbents, and (2)  the impact of technological --- product and process --- innovations on productivity.  Also, we pay particular attention to complementarity between product and process innovation for the two group of producers.	
	
Various explanations --- not limited to those proffered by innovation studies ---  for the phenomenon of declining productivity growth and job reallocation have been proposed. \citet{czarnitzki:2012} and  \citet{coad:2016} point that EU start-ups face higher entry and growth barriers, including financial constraints, than their US counterparts. \citet{akcigit:2021} argue that decline in knowledge
diffusion is likely to increase market concentration, which implies that a new entrant
is much more likely to compete against a dominant market. This, they argue, will discourage
new firm creation and is likely to result in decreased economic activity among young firms. \citet{acemoglu:2018} claim that industrial policies that subsidize (often large) incumbent firms, may also reduce economic growth by slowing down reallocation and even discouraging innovation by both continuing firms and new entrants.

\citet{acemoglu:2018}, who have a clear policy formulation for enhancing growth, build on models of endogenous technological change, models of firm-level innovation \citep{klette:2004} and incorporate major elements from the reallocation literature\footnote{This literature emphasizes selection mechanisms, which characterise industries as collections of firms that are heterogeneous in their productivity, and link firm productivity levels to their performance and survival in the industry \citep{jovanovic:1982,hopenhayn:1992,asplund:2006}. This heterogeneity induces a selection effect, by which reallocation of market shares to more efficient producers, either through market share shifts among incumbents or through entry and exit, drive aggregate productivity movements. Low productivity plants are less likely to survive and thrive than their more efficient counterparts, creating selection-driven aggregate (industry) productivity increases. \citet{decker:2014} document that, given initial size, more productive firms grow faster than the less productive ones; however, the contributions entrants make to growth are considerably heterogeneous.} to construct a model of firm innovation and growth that enables an examination of the forces that jointly drive innovation, productivity growth, and reallocation. In their model, incumbents and entrants invest in R\&D in order to improve over (one of) its many products. However, firms are heterogeneous (high and low types) in their efficiency with which they innovate. Firms that enter the market are disproportionately of high-quality type but may become low-quality firms with a certain transition probability over time. This heterogeneity introduces a selection effect, with concomitant reallocation occurring with the movement of R\&D resources (skilled workers) from less efficient innovators (struggling incumbents) towards more efficient innovators (new firms). To promote growth and welfare they, therefore, propose taxes on the continued operation of incumbents combined with a small incumbent R\&D subsidy.

\citet{aghion:2015} critiquing \citet{acemoglu:2018} point that their model, in which R\&D investments interact with general equilibrium effects, does not incorporate the notion of absorptive capacity and that it would be fruitful for future research to do so. \citet{aghion:2015}, referring to \citet{cohen:1989}, state that R\&D not only creates new knowledge but also facilitates learning and building absorptive capacity. This is particularly relevant for catching-up economies such as the Central and Eastern European (CEE) countries, which lag behind the technological frontier, and where investing in absorptive capacity through R\&D and better education can improve the ability to innovate and/or imitate leading edge technologies \citep{aghion:2010, radosevic:2019}. \citet{griffith:2004} show empirically that (a) R\&D affects both the rate of innovation and technology transfers, and therefore failing to take into account R\&D-based absorptive capacity results in large underestimates of the social rate of return to R\&D, and (b) countries and industries lagging behind the productivity frontier catch-up particularly fast if they invest heavily in R\&D. These also imply that many of the policy recommendations in \citet{acemoglu:2018} for the US and the technologically advanced economies may not be applicable for catching-up economies.     
	
\citet{parisi:2006} show that fixed capital spending by Italian medium-low and low-tech firms increases the likelihood of introducing a process innovation. This, as \citet{parisi:2006} argue, suggests that physical capital stock embodies technological progress.  In catching-up economies (e.g. Estonia) --- (a) which tend to grow more due to the imitation activities of the firms, and (b) where a higher share of firms are in medium-tech and low-tech sectors --- one could, therefore, expect capital stock accumulation to have significant productivity implications.

\citet{lubczyk:2020}(LP hereafter) state that little scholarly research has assessed the dynamics studied in \citet{acemoglu:2018} with firm level empirical applications. \citet{czarnitzki:2012} using Flemish data show that young innovative companies grow faster than the other, already fast-growing firms. \citet{coad:2016} using Spanish data show that young firms --- in terms of sales, productivity and employment growth --- benefit more from R\&D, but R\&D investment by young firms is riskier than R\&D investment by more mature firms. \citetalias{lubczyk:2020} using a novel German data set find that entrants experience significantly larger gains from investments in R\&D than incumbents and that returns to R\&D for entrants are considerably more heterogeneous than that for incumbents. Besides, \citetalias{lubczyk:2020}'s is a comprehensive study of regional spillover effects for the two groups of firms.

To assess the relevance of the above discussed theories for a catching-up economy as Estonia, the main objectives we formulate are: 
\begin{enumerate}[label=(\Roman*)]
\item  Given that reallocation, particularly due to entry and exit, accounts for a substantial proportion of productivity growth and since this reallocation, both of production and R\&D inputs, is to a significant extent due to productivity impacts of innovation undertaken by heterogeneous incumbents and entrants,  we study the productivity implications of (a) R\&D, (b) knowledge spillover, and (c) technological innovations for incumbents and entrants. 

\item  Since catching-up economies have been known to grow by engaging in imitation/learning activities  \citep[][]{aghion:2010, radosevic:2019}, embodied technological change through capital accumulation is likely to have a significant productivity impact. We, therefore, compare the productivity impact of capital accumulation with the productivity impact of  R\&D for both groups of producers.

\item  Our third objective is to compare the productivity growth impacts of R\&D and capital stock for Estonia, a catching-up country, to those in \citetalias{lubczyk:2020}'s, who have undertaken a similar analysis for Germany, a more technologically advanced country, and understand the sources of the difference in the impacts. \citetalias{lubczyk:2020}'s, to our knowledge, is the most recent work that has comprehensively looked at productivity impact of R\&D and regional spillovers for the two groups of producers in a developed economy. While the comparison puts our results in perspective, we believe it will also inform better innovation and growth enhancing policies for Estonia and catching-up economies. 
\end{enumerate}

	%, to compare entrants and incumbents with regards to how much they benefit from their own R\&D in terms of productivity improvements. We also look at how much these two group of firms, young and old, learn and benefit from R\&D done by other firms in proximity. 

Before proceeding further, we document some of our findings. To begin with, though all firms gain from investing in R\&D, we do not find any difference in the average productivity elasticity with respect to R\&D for the incumbents and the entrants. Our results therefore do not match the findings in \citet{coad:2016}, \citet{acemoglu:2018} and \citetalias{lubczyk:2020}, where the impact of  R\&D on productivity is much higher for entrants as compared to the incumbents'. However, we do find that impact of technological innovation --- many of which are a result of DUI mode of innovation --- on productivity is much higher for the entrants as compared to the incumbents. We explored the sources of the productivity gain for both the groups of producers, and found that the gain is largely due to complementarity between product and process innovation; innovating entrants, however, realize a much higher complementarity than incumbents.

Now, while Estonian entrants were found to be less able than their counterparts in Germany in translating R\&D into productivity gains, we found an average Estonian incumbent to be on par with an average German incumbent. This is despite the adverse sectoral composition of the Estonian economy, where, as is typical of catching-up CEE economies, compared to an average European Union member state, there are fewer firms and fewer people employed in sectors where the returns to R\&D are highest. Moreover, because the bulk of STI based innovative activities in Estonia is carried out by the incumbents, we find  Estonian incumbents to be playing an important role as carriers of STI based innovative activities.

We find a robust evidence that embodied technological change through capital accumulation is more effective in generating productivity growth than R\&D expenditure. However, since R\&D or STI based innovative activities itself help build absorptive capacity, we also find that productivity elasticity with respect to capital is higher for firms that report engaging in formal R\&D. Finally, in results related to spillover effects, we find it is mostly the incumbents who benefit from intra- and inter-industry knowledge that is generated by other firms --- primarily by the incumbents --- situated in close proximity. However, the spillover effects are higher for firms that do not reportedly engage in formal R\&D, suggesting that these firms, too, work on building absorptive capacity.

The remainder of this paper is structured as follows. Section 2 reviews related literature and in section 3 we describe the data used for our study. In section 4, we explain the empirical strategy employed in the paper. In Section 5, we present and discuss the empirical results, while section 6 draws concluding remarks. Certain details of the econometric methodology have be relegated to the appendix.

	\section{Literature Review}
	In subsection 2.1 we discuss some literature on the relationship between firm age, innovation and productivity, while in subsection 2.2 we review some relevant studies on innovation, reallocation, and productivity growth for the CEE countries.
	
	\subsection{Innovation among Entrants and Incumbents}
	
	Among the large strand of literature on how different firms contribute to productivity and productivity growth, many studies have examined how dynamics between entrants and incumbents impact aggregate productivity development, highlighting both differential and interrelated effects. In this section, we review some of the literature on the innovative behaviour of the two groups of firms. 
	
	Entrants and incumbents have been described as two distinct groups of firms \citep{berchicci:2009, lubczyk:2020}. Entrants wanting in experience, still need to learn about the economic environment in which they operate. Incumbents, on the other hand, have accumulated considerable experience in their competitive environment and command well established capabilities. Entrants, however, are seen as being more expeditious than incumbents due to lack of structural inertia to reorganisation, faster decision-making processes, streamlined operations,  and targeted innovation. These result in a timely response to changing industry environments, and also make them more efficient innovators compared to incumbents.
	
	However, entrants are often financially constrained, and investing a large part of the limited resource endowment in R\&D activities, which are inherently uncertain, can pose a significant business risk. Lack of experience and limited expertise may create further complications for the innovation process, especially when unforeseen circumstances arise. On the other hand, however, entry is envisaged as the way in which firms explore the value of new ideas in an uncertain context, and that entry, the likelihood of survival, and subsequent conditional growth are determined by barriers to survival \citep{audretsch:1995,Huergo:2004}. In this framework, entry is innovative and occurs at a higher rate at the start of an industry  when uncertainty is high \citep{klepper:1996}; the likelihood of survival is lower the higher the risk; and the growth from successful innovation is higher the higher the barriers to survival. Successful innovation, therefore, is likely to result in substantial relative productivity growth for newly established enterprises. Given that most young firms are small, successful innovation can often disproportionately spur growth and contribute to substantial increases in employment, revenue and future profitability \citep{haltiwanger:2013}. Therefore, one would expect the impact of R\&D spending on productivity to be volatile for the new entrants.  
	
	Incumbents, having survived through time and participated longer in the market, and have many factors working in their favour as far as innovative activity is concerned. First, even though incumbents often lack the organisational agility of smaller and younger competitors, they may compensate for this with resources --- such as financial and marketing capabilities ---  and innovation capacity built over time \citep{berchicci:2009}. Moreover, an incumbent can build on its existing infrastructure even as existing business experience and infrastructure may enable the incumbent to pursue more ambitious R\&D projects. Also, the experience of having conducted successful innovation in the past increases the likelihood of future innovation \citep{peters:2009, raymond:2010} and may help such organisations to achieve higher levels of efficiency in carrying out their R\&D activities \citep{loof:2014}. 
	
	The nature of innovations, too, is often different for the two groups of producers. Since incumbents would like to safeguard their profits from the established products and production technologies in place, their innovative activity is therefore more often of an incremental nature, whereas young firms, in order to create higher quality products and overtake product lines previously operated by incumbents, are more inclined to exploit new ideas and engage in radical innovation \citep{acemoglu:2015}. Besides, radical innovation often entails costly organisational restructuring, which may deter incumbents from undertaking radical innovation \citep{berchicci:2009}.
	
	\subsection{Relevant Studies on Innovation and Reallocation in CEE Countries }
	
	While there are many studies on innovation, reallocation, and productivity growth for the US and Western European countries, there are only a few as far as Central and East European (CEE) countries are concerned. Studies on productivity growth due to selection and reallocation for the CEE countries are by \citet{masso:2004} and \citet{bartelsman:2013}. \citet{masso:2004} find that newly formed firms had a higher survival rate than incumbents, and that the reallocation of production factors, especially due to the exit of low productivity units, contributed to the productivity growth. \citet{bartelsman:2013} find that, although the covariance between firm-size and productivity, a measure of resource misallocation, is low in Eastern Europe, it has been increasing substantially over the last couple of decades.

	\citet{masso:2008} point out that to sustain initial growth rates during the transition period in CEE countries, which was based on initial capital accumulation and imitation of technologies applied elsewhere, these countries will need to rely increasingly on their own innovation. Due to their attempts to establish knowledge-based economies and to increase business R\&D, there is growing interest in studying the relationship between innovation, productivity and growth in the CEE countries. Among the few studies that have studied the productivity impact of R\&D in a CEE country is one by \citet{liik:2014}, who estimate elasticities using industry level data to conclude that in comparison to OECD economies, R\&D investments play a relatively limited role in determining the productivity and efficiency levels of Estonian industries. \citet{lacasa:2017}, studying the technological capabilities of CEE economies based on patent data, find that the CEE economies reduced their technological activities drastically after 1990, and that the recovery of CEE economies with respect to technological capabilities is unfolding very slowly. They find that CEE countries innovate in less dynamic technological sectors and contribute only to a limited number of fields with growing technological opportunities.  
	
	Recent studies, such as \citet{bruno:2019}, find that while R\&D intensity has been effective in closing the distance to the productivity frontier, R\&D embedded in purchased equipment and machinery have played an important role in reducing the distance. \citet{filippetti:2015}, studying the role of the technology gap in explaining labour productivity differences in 211 European regions over the years 1995---2007, find that labour productivity growth in CEE (or lagging behind) regions, productivity growth is mainly driven by capital accumulation. 
	
	\section{Data and Variables: Definitions and Description}
	
	For our study, we use eight waves of the Estonian Community Innovation Survey (CIS), CIS2004 to CIS2018, which is a biennial survey about innovation activities in Estonian enterprises. For a detailed description of the CIS data and the usage of the data for scholarly analysis of various issue related to innovation, see \citet{mairesse:2010}. Firm balance sheet information containing profit and loss statements was obtained from the Estonian Business Registry, which is a census data. Wherever possible, the missing information in the Business Registry data was obtained from EKOMAR, a survey data.

	One of the key variables from the CIS surveys used for our analysis is the R\&D expenditure of firms. However, as non-innovative firms --- firms that have not innovated any product or process, or have no unfinished innovative activities --- in the CIS data are not required to report their R\&D expenses, we do not observe R\&D expenditure for the non-innovative firms. The CIS questionnaire, however, allows us to distinguish between `potentially' innovative firms and firms that do not intend to engage in R\&D  among the non-innovative firms \citep[see][]{savignac:2008}. This distinction is made on the basis of firm responses to the question on factors that might have thwarted their R\&D and innovative activities. We classify non-innovative firms who face such factors as potentially innovative, whereas non-innovating firms that do not face any of these factors are classified as firms that do not intend to engage in R\&D activities. For these non-innovators that do not wish to engage in R\&D activities, it can be safely assumed that they have no R\&D expenses.   
	
	Since we do not observe R\&D expenditure of potentially innovative firms, we drop such firms from our analysis. However, non-innovative firms that do not wish to engage in R\&D activities are retained. This, as discussed in the next section, allows for us to estimate the spillover effects of external knowledge stock as well as the impact of product and process innovation on productivity for all firms: firms that have and that do not have R\&D expenses. After further removing firms with missing data, our sample comprised of 7,586  firm-year observations from 3,068 firms. The minimum number of observations per firm is 1, the maximum, 8, and the average number is about 2.5 years. 
	
	Since one of our aims is to compare our findings to those in \citetalias{lubczyk:2020}, we, as in \citetalias{lubczyk:2020}, define entrants as firms that are new to the market and have been active for eight or less years, while incumbents are those that have been active for more than eight years. Our set of entrants excludes firms resulting from  mergers, break-ups, split-off or restructuring of a set of enterprises. Those resulting from change of activity, too, are excluded. There is no theoretical basis for the above definition of an entrant; in the literature, an entrant's maximum age has ranged from three to ten years. Of the 7,586 firm-year observations, 6,086 are incumbents and 1,500 are entrants. Estonian firms are relatively young, where the average age of incumbents is 17 years, while that of entrants is 5.5 years.

	%Birth amounts to the creation of a combination of production factors with the restriction that noother enterprises are involved in the event. Births do not include entries into the population due to: mergers, break-ups, split-off or restructuring of a set of enterprises. It does not include entries into a sub-population resulting from change of activity. Only companies that have begun their economic activity in a given year are counted as births.latest data published for 2011. Maximum birth date given. If there is no birth marked in the table, the enterprise was not in the Business Demography population by the above explained methodology; in that case, registration date could be used.

	Also, we define a firm, $ j $,  in year, $ t $, to be an innovator, $ I_{jt}= 1 $, if the firm had introduced new products, $ \mathcal{P}^{D}_{jt}= 1 $ or processes,  $ \mathcal{P}^{C}_{jt}= 1 $, to the market. Value-added productivity is our dependent variable and performance outcome measure. It is the ratio of value-added, the difference between sales revenues and the value of intermediate inputs, to the number of employees.
	
	In order to convert the book value of the gross capital stock into its replacement value, we use the perpetual inventory method described in \citet{salinger:1983}. According to this method, the replacement value of the capital stock is equal to the book value of fixed assets for the first year the firm appears in the data. For the subsequent years, first, the useful life of capital goods, $ L_{jt} $, at time $ t $ is calculated as:
	\begin{align}\nonumber
	L_{jt} = \frac{GK_{j,t-1} + I_{jt}}{ DEPR_{jt}}, 
	\end{align}
	where $GK_{j,t-1}$ is the reported value of gross property, plant, and equipment at time $ t - 1 $, $ I_{jt} $ is the investment in the same for the period $ t $, and $ DEPR_{jt} $ is the reported depreciation. Then  $ L_{jt} $ is averaged over time to obtain $ L $, which is then used in the following formula to obtain the replacement value of the capital stock of a firm in industry $ k $:
	
	\begin{align}\nonumber
	K_{jt} =\biggr(K_{j,t-1}\dfrac{p^{k}_{t}}{p^{k}_{t-1}} + I_{jt}\biggr) (1 - 2/L_{jt}), 
	\end{align}
	where $ p^{k}_{t} $ is the deflator for industry, $ k $.  The second term represents the amount of capital stock that depreciates each year and is based on the assumption that economic depreciation is a double declining balance. For new firms and for existing firms that appear again after a gap in later time periods in the data, the book value of the capital stock in the first year is taken as the replacement value. However, for firms that after a gap reappear in later years, this method will not yield as good an estimate of the replacement value as for the firms for whom a long, continuous time-series is available.

	In Table \ref{table:ds2}, we can see that the incumbents on average have significantly higher labour productivity, larger capital stocks, higher investment and more employees. As far as innovative behaviour of the firms is concerned, as can be seen in Table \ref{table:ds2}, a majority of the firms (about 74\%) in the estimation sample are innovative even as the percentage of firms that have positive R\&D is only about 30\%. Secondly, incumbents have a higher propensity to invest in R\&D and are also likely to invest more in R\&D than the entrants. However, it is the entrants who are more innovative as far as technological --- product and process --- innovations are concerned. Also, for the incumbents, log of R\&D intensity increased substantially from an average of 1.76 in the pre Financial Crisis years to an average 6.93 in the post Crisis years; the corresponding figures for the entrants are, 2.03 and 6.95.
	
	In Table \ref{table:ds3}, we can see that high-tech and medium-high-tech manufacturing, which are most R\&D intensive and most innovative, constitute about 13\% of the firms in the sample. Majority of the firms (70\%) are in the low-tech and medium-low-tech manufacturing and less knowledge intensive services; these sectors, though less R\&D intensive, are still quite innovative. 
	%This also suggests that by removing `potentially innovative' firms (firms that could have innovated but did not) from our sample, our sample is biased towards firms that have innovated.   
	
	In our analysis we include a dummy for north Estonia, which takes value 1 if the company is located in Harju county. Harju county is the biggest of all Estonian counties in terms of population and economic activity, and includes the national capital city, Tallinn. The rational for including this dummy is that firms located in Harju county, which qualifies as the economic hub of Estonia, are likely to be better networked with implications for productivity due to agglomeration. We find that a significantly higher proportion of entrants are based in northern Estonia (see Table \ref{table:ds2}).  
	
	The external knowledge capital, $ E_{jt} $, is intended to capture knowledge spillovers among firms. In our empirical analysis, we differentiate between different types of spillovers: intra-industry  and inter-industry  R\&D spillovers to measure to what extent entrants and incumbents differ in their capacity to benefit from R\&D knowledge that is available to the firm within and outside its own industry.
	
	The measure of the intra-industry knowledge capital of firm $ i $ in industry $ k $ (based on two digit NACE Rev. 2 codes)  in period $ t $ is the weighted sum of R\&D expenditures per employee of other firms in industry $ k $: 
	\begin{align}\label{eq:1}
	\text{ Intra-Industry R\&D } = \sum_{j\neq i}w_{ij} \frac{R^{k}_{jt}}{L^{k}_{jt}},
	\end{align}
	where $ R_{jt} $ is the R\&D expenditure of firm $ j $, $ L_{jt} $ is the number of employees employed by the firm, and the weight, $ w_{ij} $, is the inverse of the geographical distance between the capital of the county in which firm $ i $ is located and capital of the county in which firm $ j $ belonging to the same industry, $ k $, is located; if firm $ j $ happens to be in the same county then $ w_{ij} $ is taken as 1. The measure of the inter-industry knowledge of firm $ i $ in industry $ k $ (NACE two digit) in period $ t $ is the weighted sum of R\&D expenditures per employee of firms in other industries:
	\begin{align}\label{eq:2}
	\text{ Inter-Industry R\&D } = \sum_{l\neq k} \sum_{j} w_{ij}\frac{R^{l}_{jt}}{L^{l}_{jt}}, 
	\end{align} where, again, the weight $ w_{ij} $ is the inverse of the geographical distance between the capital of the county in which firm $ i $ is located and the capital of the county in which firm $ j $ belonging to a different industry, $ l $, is located.

	In Table \ref{table:ds2} we can see that incumbents on average are based in or based close to regions/counties with higher average external knowledge; the differences hold for both within and across industry comparisons.  Here, we would like to note that even at this broader definition of industry (NACE 2-digit), about 2\% of the firm-year observations had zero intra-industry knowledge flows. This is because a relatively smaller number of firms, about 30\%, in the estimation sample invested in R\&D.

	\begin{center}
		\textbf{[Table \ref{table:ds2} about here]}
	\end{center}
	
	\begin{center}
		\textbf{[Table \ref{table:ds3} about here]}
	\end{center}

	\section{Empirical Strategy}\label{sec4}

	For firm $ j = 1, \ldots , J $ and time $  t = 1, \ldots , T $, we observe revenue ($ Y_{jt} $), labour ($ L_{jt} $), capital ($ K_{jt} $), material inputs ($ M_{jt} $),  R\&D expenses ($ R_{jt} $), which can also be zero, and ($ E_{jt} $), which is some measure of external knowledge capital. 
	
	Now, while a measure of R\&D capital stock would be preferred as an innovation input, we are unable to estimate the knowledge capital from the past R\&D expenditures using the perpetual inventory method. This is because CIS surveys are conducted once every two years and, secondly, we do not have a balanced panel data. Instead, following \cite{cerpon:1998} and more recently \citet{raymond:2015} and \citet{baumann:2016}, we proxy knowledge capital using current R\&D expenditure, $ R_{jt} $. This proxy is based on the implicit assumption that firm R\&D investments are strongly correlated (and roughly proportional) to their R\&D capital stock measure, and that R\&D engagement and intensity persists over time.\footnote{With a big overlap of firms between the different waves in the Estonian CIS, we find some evidence of this persistence.}

	Let $ \omega_{jt} $ be the firm $ j $ and time $ t $ specific total factor productivity (TFP), which is observed by the firms but not by the econometricians. For notational convenience, we drop the subscript, $ jt $, as of now. For estimation, we employ the ``control function" methodology developed by \citet{ackerberg:2015} (ACF), who consider a value-added production function. The value-added production function, as \citetalias{ackerberg:2015} explain, can be derive from the gross output production function that is Leontief in material inputs, $ M $. So, when R\&D expenditure is positive, i.e. $ R>0 $, output is given by  
	\begin{align}\nonumber
	Y &= \min\{B L^{\beta_{l}}K^{\beta_{k}}R^{\beta_{r}}E^{\beta_{e}}e^{\omega}, \beta_{m}M \}.
	\end{align}
	The above implies that the value-added productivity, $ \dfrac{Y -M}{L} $, when  $ R>0 $  can be written as 
	\begin{align}\label{eq:3}
	\dfrac{Y -M}{L} = (1-\frac{1}{\beta_{m}})B L^{\beta_{l}+ \beta_{k}+\beta_{r}-1}\biggr(\dfrac{K}{L}\biggr)^{\beta_{k}}\biggr(\dfrac{R}{L}\biggr)^{\beta_{r}}E^{\beta_{e}}e^{\omega}.  
	\end{align}
	
	As discussed in the section on data, not all firms, innovating or otherwise, have positive R\&D expenditure. We nonetheless keep such firms in our analysis for two reasons. First, this allows us to estimate the spillover effects of external knowledge capital for firms that do not engage in R\&D. Second, as discussed below, we are able to use the information on product and/or process innovation, even for firms that do not have positive R\&D expenses, to endogenise the evolution of $ \omega $ and assess the implications of innovation on productivity, $ \omega $.
	The Leontief production function when R\&D expenditure, $ R $, is equal to zero and when material input is proportional to output is given by 
	\begin{align}\nonumber
	Y &= \min\{B_{0} L^{\beta^{0}_{l}}K^{\beta^{0}_{k}}E^{\beta^{0}_{e}}e^{\omega}, \beta^{0}_{m}M^{} \},
	\end{align}
	which implies that the value-added productivity when $ R=0 $ is given by
	\begin{align}\label{eq:4}
	\dfrac{Y -M}{L} = (1-\frac{1}{\beta^{0}_{m}})B_{0}L^{\beta^{0}_{l}+\beta^{0}_{k}-1}\biggr(\dfrac{K}{L}\biggr)^{\beta^{0}_{k}}E^{\beta^{0}_{e}}e^{\omega}.
	\end{align}
	Taking logarithm of the value-added production functions when $ R>0 $ and when $ R=0 $ and combining the two we get 
	\begin{align}\label{eq:5}
	y = &  (1- D^{0})(\beta+ \beta^{+}_{l}l + \beta_{k}k  + \beta_{r}r +\beta_{e}e)   + D^{0}(\beta^{0} + \beta^{0+}_{l}l + \beta^{0}_{k}k +\beta^{0}_{e}e )  + \omega+\epsilon, 
	\end{align}    	
	where $ y $ is the natural logarithm of the value-added productivity and $ D^{0} $ is a dummy variable that takes the value 1 when R\&D expenditure is zero. The term $ \beta^{+}_{l} = \beta_{l}+ \beta_{k} + \beta_{r} - 1 $ and $ \beta^{+0}_{l} = \beta^{0}_{l}+ \beta^{0}_{k} - 1 $. The lower-case symbols, $l$, $k$, $r$, and $ e $, on the RHS represent natural logs of $ L$,  $\dfrac{K}{L} $,   $ \dfrac{R}{L} $, and $ E $ respectively. The constants $ \beta $  and $ \beta^{0} $ are $\ln(B)$ and  $\ln(B_{0})$ respectively. The term $ \epsilon $ is the measurement error in value-added productivity.\footnote{See \citet{ackerberg:2015} for a discussion on the interpretation of $ \epsilon_{jt} $.} Since the specification in (\ref{eq:5}) log-linear, the coefficients of the input variables can be interpreted as elasticities. As is evident from the equation, we allow the productivity elasticities with respect to the input variables to be different for firms that engage in R\&D and firms that do not.

	Since the unobserved productivity shocks, $\omega$, are known to the firm when it makes its input choices, we are confronted with the problem of endogeneity.  To resolve the issue of endogeneity, as stated earlier, we employ the `control function' method by \citetalias{ackerberg:2015}. The method  requires that the outcome variable be value added, and depending on the timing of investment and input decisions, it involves the use of economic theory to derive a `proxy' for the anticipated productivity shock, $\omega$. The proxy is obtained by assuming that $\omega$ can be inverted out from certain firm inputs if the firm has adjusted these inputs optimally in response to the $\omega$ it observed.

	Now, in \citetalias{ackerberg:2015} the evolution of productivity, $ \omega_{jt} $, has been assumed to be exogenous; that is, productivity shocks evolve according to the first order Markov Process: $ p(\omega_{j,t+1}|\mathcal{I}_{jt}) = p(\omega_{j,t+1}|\omega_{jt}) $, where $\mathcal{I}_{jt}$  firm's information set at $t$, which includes current and past productivity shocks $\{\omega_{j\tau} \}^{t}_{\tau=0} $ but does not include future productivity shocks $\{\omega_{j\tau}\}^{\infty}_{\tau=t+1}$. The distribution, $  p(\omega_{j,t+1}|\omega_{jt}) $, which is stochastically increasing in $ \omega_{jt} $, is known to the firm.\footnote{If only to mention, the transitory shocks, $ \epsilon_{jt} $, in (\ref{eq:5}) satisfy $ \E[\epsilon_{jt}|\mathcal{I}_{jt}] = 0 $.}
	
	\cite{griliches:1979}, however, points out that while R\&D affects output, R\&D is determined both by past output and the expectations of future output. And thus past R\&D efforts and innovation output affect the evolution of productivity, $\omega_{jt}$,  and the expectation of $ \omega_{jt+1} $ affects the endogenous choice of $R_{jt}$. \citet{doraszelski:2013}(DJ) endogenize productivity evolution by including a measure of R\&D investment: $ p(\omega_{j,t+1} | \omega_{jt}, R_{j,t}) $. According to this formulation, a firm by making investments in R\&D alters the probability of receiving a innovation, which in turn alters the distribution of productivity that it faces in future periods.
	
	\citet{peters1:2017}, endogenize productivity evolution by making $\omega_{jt} $ depend on innovation output instead of past innovation input. More specifically, productivity, $ \omega_{jt} $, is modelled to evolve according to the following controlled Markov process: $ p(\omega_{jt} | \omega_{j,t-1}, \mathcal{P}^{D}_{j,t}, \mathcal{P}^{C}_{j,t})$, where $ \mathcal{P}^{D}_{j,t} $ is an indicator variable that takes the value 1 if firm $ j $ innovated at least one product and 0 otherwise, and $ \mathcal{P}^{C}_{j,t} $ is an indicator variable that takes the value 1 if firm $ j $ innovated new processes in period $ t $ and zero otherwise. Here the assumption about the innovation process is that both $ \mathcal{P}^{D}_{j,t} $ and $ \mathcal{P}^{C}_{j,t} $ are a result of R\&D efforts in the past. By including the innovation process in the model, rather than linking R\&D directly to productivity as in \citetalias{doraszelski:2013}, \citet{peters1:2017} aim to gain additional insight into whether R\&D improves productivity through the demand side or the cost side of the firm's operations. This is because product innovations increase revenue by expanding the firm's demand, whereas process innovations increase revenue by improving the efficiency of scarce  factors and thereby reducing firm's cost of production \citep[see also][]{hall:2011}. 
	
	The Markovian assumption implies that
	\begin{align}\label{eq:9}
	\omega_{jt} &= \E(\omega_{jt}|\omega_{j,t-1},\mathcal{P}^{D}_{j,t}, \mathcal{P}^{C}_{j,t}) + \xi_{jt}=  g(\omega_{j,t-1}, \mathcal{P}^{D}_{j,t}, \mathcal{P}^{C}_{j,t}) + \xi_{jt}. 
	\end{align}
	In the above, productivity $ \omega_{jt} $ in period $ t $ has been decomposed into expected productivity, $ g(\omega_{j,t-1}, \mathcal{P}^{D}_{j,t}, \mathcal{P}^{C}_{j,t}) $, and a random shock, $ \xi_{jt} $. While the conditional expectation function $ g(.)$ depends on the already attained productivity $ \omega_{j,t-1} $ and $ (\mathcal{P}^{D}_{j,t}, \mathcal{P}^{C}_{j,t}) $, $ \xi_{jt} $ does not. The residual, $\xi_{jt}$, by construction is mean independent of $ \omega_{j,t-1} $ and $ (\mathcal{P}^{D}_{j,t}, \mathcal{P}^{C}_{j,t}) $. As discussed in  \citetalias{doraszelski:2013}, the productivity innovation, $ \xi_{jt} $, represents the uncertainties that are naturally linked to productivity and the uncertainties inherent in the innovation process such as degree of applicability and success in implementation. 
	
	The formulation of productivity evolution in (\ref{eq:9}) (a) captures the lag dependence of productivity, $ \omega_{jt} $, and (b) by allowing innovations to affect the expected productivity, $ g(.) $, and hence labour productivity, it captures the fact that R\&D expenditures alone are not sufficient to generate productivity improvements. This formulation, therefore, allows us to estimate the productivity impact of technological innovations. The conditional expectation function $ g(.) $ for the specification in  (\ref{eq:5}) is estimated non-parametrically along with the parameters of the production using the two step control function procedure proposed in \citetalias{ackerberg:2015}.\footnote{We use the STATA's  `prodest' command, which has been developed by \citet{rovigatti:2018}, to implement the control function method in \citetalias{ackerberg:2015}. The endogenous() option allows users to specify one or more variables that endogenously affect the dynamics of productivity, $ \omega_{jt} $. }

	Since we are interested in estimating the impact of technological innovations on productivity for incumbents and entrants, we estimate the difference,   
	\begin{align}\label{eq:10}
	\Delta=\E[g_{1 \vee 1}(\omega_{j,t-1}, .)] - \E[g_{0\wedge 0}(\omega_{j,t-1}, .)]
	\end{align}    
	for the entrants and the incumbents, where
	\begin{align}\nonumber
	& g_{1 \vee 1}(.) \equiv g(\omega_{j,t-1}, \mathcal{P}^{D}_{jt} , \mathcal{P}_{jt}^{C}) \text{ such that } \mathcal{P}^{D}_{jt} =1  \textit{ and/or } \mathcal{P}^{C}_{jt}=1 \text{ and }\\ \label{eq:11}
	& g_{0\wedge 0}(.) \equiv g(\omega_{j,t-1}, \mathcal{P}^{D}_{jt} , \mathcal{P}_{jt}^{C}) \text{ such that } \mathcal{P}^{D}_{jt} =0  \textit{ and } \mathcal{P}^{C}_{jt}=0, 
	\end{align}
	and the expectations are taken over $ \omega_{j,t-1} $. The difference, $ \Delta $,  in (\ref{eq:10}) is the average treatment effect (ATE) of innovation on productivity; details on the identification of ATE can be found in Appendix \ref{ap_a}.

	To understand how (i) product innovation only, (ii) process innovation only, and (iii) product \textit{and} process innovations contribute to the ATE, $ \Delta $, of technological innovation we estimate the following:
	\begin{align}\nonumber
	&\Delta_{10}=\E[g_{1\wedge 0}(.)] - \E[g_{0\wedge 0}(.)],\\\nonumber
	&\Delta_{01}=\E[g_{0 \wedge 1}(.)] - \E[g_{0\wedge 0}(.)],  \text{ and }  \\\label{eq:12}
	&\Delta_{11}=\E[g_{1\wedge 1}(.)] - \E[g_{0\wedge 0}(.)],
	\end{align} 
	where 
	\begin{align}\nonumber
	& g_{1 \wedge 0}(.) \equiv g(\omega_{j,t-1}, \mathcal{P}^{D}_{jt} , \mathcal{P}_{jt}^{C}) \text{ such that } \mathcal{P}^{D}_{jt} =1  \textit{ and  } \mathcal{P}^{C}_{jt}=0, \\\nonumber 
	& g_{0 \wedge 1}(.) \equiv g(\omega_{j,t-1}, \mathcal{P}^{D}_{jt} , \mathcal{P}_{jt}^{C}) \text{ such that } \mathcal{P}^{D}_{jt} =0  \textit{ and } \mathcal{P}^{C}_{jt}=1, \text{ and } \label{eq:13} \\
	& g_{1 \wedge 1}(.) \equiv g(\omega_{j,t-1}, \mathcal{P}^{D}_{jt} , \mathcal{P}_{jt}^{C}) \text{ such that } \mathcal{P}^{D}_{jt} =1  \textit{ and } \mathcal{P}^{C}_{jt}=1. 
	\end{align}
	Given the definitions of $ g_{1 \wedge 0}(.) $, $ g_{1\wedge 0}(.) $,  and $ g_{1\wedge 0}(.) $ in (\ref{eq:13}) and $ g_{0 \wedge 0}(.) $ in (\ref{eq:11}), it is clear that $ \Delta_{10} $ in (\ref{eq:12}),  for the matched sample of product innovators only and non-innovators, is the ATE of product innovation only; $ \Delta_{01} $, the ATE of process innovation only; and  $ \Delta_{11} $, the ATE of product \textit{and} process innovation.\footnote{Ideally, $ \Delta $ in (\ref{eq:10}) is the weighted sum, $ \Delta = w_{10}\Delta_{10} + w_{01}\Delta_{01} + w_{11}\Delta_{11} $. The weights, $  w_{10}$, $ w_{01}$, and $ w_{11} $, respectively are $  \dfrac{N_{10}}{N} $, $ \dfrac{N_{01}}{N}$, and $ \dfrac{N_{11}}{N} $, where $ N_{10} $ is the number of firms that innovated only products, $ N_{01} $ is the number of firms that innovated only processes, $ N_{11} $ is the number of firms that innovated both products and processes, and $ N =  N_{10} +  N_{01} + N_{11}  $. However, if, for example, we use different matching criteria for computing $ \Delta_{11} $ and $ \Delta $ so that the sample of firms used for computing $ \Delta_{11} $ is not contained in the sample of firms used to compute $ \Delta $, then the equality may not hold.} As shown in Appendix \ref{ap_a}, if 
	\begin{align}\label{eq:14}
	\Delta_{11} \geq \Delta_{10} + \Delta_{01},
	\end{align}    
	that is, if the impact of product \textit{and} process innovation exceeds the sum of the impacts of \textit{only} product innovation and \textit{only} process process innovation, then the product and process innovation can be said to be complementary. We refer to \citet{milgrom:1990} for a general treatment of complementarities between organizational decisions, and to \citet{athey:1995} for complementarities between product and process innovation.

	\section{Results}
	
	First, in subsection \ref{sec51}, we discuss the results obtained from the estimation of the production functions in equations (\ref{eq:5}), focusing on the average impacts of internal R\&D investments and capital accumulation on productivity for the entrants and incumbents. In this subsection, we also  compare our estimates of the impact of internal R\&D and capital accumulation on productivity with those estimated by \citet{lubczyk:2020}(LP) for Germany. In subsection \ref{sec52} we study the differential spillover effects from knowledge that is produced outside the firms' boundaries. Finally, in subsection \ref{sec53}, we discuss the estimates of the impact of technological innovation on productivity for the two group of producers.    
	
	\subsection{Average Returns to Own R\&D and Capital Accumulation for Entrants and Incumbents} \label{sec51}
	
	Using the empirical framework outlined in section \ref{sec4}, we first study how the productivity effects of (1) innovation effort as proxied by R\&D expenditure and (2) capital accumulation differ between entrants and incumbent firms. In Table \ref{tabres1}, we display the results of estimating equation (\ref{eq:5}). 
	
	In Table \ref{tabres3} we compare the estimated elasticities of productivity with respect to own R\&D and capital stock with those estimated in \citetalias{lubczyk:2020} for Germany.  An important difference when compared to Germany is that in Estonia, a large percentage of innovating firms do not engage in formal R\&D. Estimating only the productivity elasticity with respect to R\&D would, therefore, give a partial picture of productivity implication of innovation. We, therefore, in addition to estimating productivity  elasticity with respect to R\&D, as discussed earlier, in subsection \ref{sec53} estimate the impact of technological innovations on productivity along. This required us to keep all firms in the estimation sample, and since the productivity elasticities with respect to other inputs such as labour and capital could well differ for firms with positive R\&D (firms that engage in STI based innovative activities) and zero R\&D, we estimated separate productivity elasticities for firms with positive R\&D and firms with zero R\&D.
	
	We find that the average productivity elasticity of R\&D expenditures is significantly positive for both entrants and incumbents.  However, entrants' elasticity of productivity with respect to R\&D does not differ from that of incumbents' systemically across specifications and methods. This is contrary to the findings in \citet{acemoglu:2018} and \citetalias{lubczyk:2020}, where entrants are disproportionally abler than incumbents in translating their R\&D endeavours into productivity gains.

	Now, in our sample there are fewer firms in sectors that are R\&D and knowledge intensive, and where returns to R\&D is high. In our sample, about 30\% of the firm-years have positive R\&D, which is mostly concentrated in high-tech and medium-high-tech manufacturing (see Table \ref{table:ds3}). However, in terms of value-added, we find that the average (averaged over the years) share of high-tech manufacturing within manufacturing is less than 7\%, and average share of medium-high-tech manufacturing is about 22\%. Though the shares have generally increased over the year, they are lower than the corresponding EU27 2006 shares of 12.5\% and 32\% respectively \citep[see][chapter 5, p. 121]{eu:2011}. Therefore, it could be due what is known as the compositional effect \citep[see][]{castellani:2019} that our result is not consistent with the theoretical argument that young firms that do not have a well established product portfolio on the market benefit more from investing in R\&D.   
	
	To examine whether, compared to the technologically advanced countries, the lower average elasticity with respect to R\&D for Estonian entrants is due to the compositional effect, in Table \ref{tabres4} we estimated separate elasticities for (a) firms in the high-tech and medium-high-tech manufacturing and (b) firms in the remaining sectors. We find that entrants in high-tech and medium-high-tech manufacturing are not particularly abler than entrants in other sectors in translating R\&D into productivity gains. Having ruled out compositional effect as a likely cause for the lower average elasticity with respect to R\&D, we take it as given that compared to their counterparts in technologically advanced economies, Estonian entrants' are inefficient in STI mode of innovative activities. Further exploration of the causes behind the comparatively lower elasticity with respect to R\&D is outside the scope of the paper.

	Incumbents in high-tech and medium-high-tech manufacturing, as we can see in Table \ref{tabres4}, do earn a significantly higher reward for investing a unit more of R\&D than incumbents in other sectors. In other words, had the sectoral composition of the Estonian economy been similar to the technologically advanced countries or to the EU average, the average elasticity with respect to R\&D for Estonian incumbents would have been higher.   
	
	So, (a) despite the limitation posed by the sectoral composition, as can be seen in Tables \ref{tabres1} and \ref{tabres3},  Estonian incumbents are no less abler than Estonia entrants and on par with German incumbents in translating R\&D efforts into productivity gains. Now, unlike in \citetalias{lubczyk:2020}'s paper for Germany, where more than half of the firms are entrants, and all of whom engage in formal R\&D, in our sample, about 20\% of the firms are entrants,\footnote{Since CIS considers only those firms which have more than ten employees and since smaller firms are likely to be younger, our sample might be biased against the entrants. However, even if we were to consider all the young firms, it is unlikely that the number of entrants would have increased substantially. Also, it would be unlikely that among the entrants left out, a higher percentage of them would be engaging in formal R\&D.} of which only 30\% are engaged in formal R\&D. In other words, (b) the bulk of STI based innovation in Estonia is carried out by Estonian incumbents. Given (a) and (b), we can, therefore, conclude that in Estonia, incumbents play an important role as carriers of STI mode of innovation. STI mode of innovation, which involves highly skilled scientific personnel, exploits possibilities at the frontier of knowledge and, through innovations, pushes the frontier, too. Besides, also through it's interconnections with the DUI mode, the STI mode enhances the ability of economies to import and adapt new technologies and innovations developed elsewhere \citep{jensen:2007}.

	\begin{center}
		\textbf{[Table \ref{tabres1} about here]}
	\end{center}
	
	\begin{center}
		\textbf{[Table \ref{tabres3} about here]}
	\end{center}
	
	\begin{center}
		\textbf{[Table \ref{tabres4} about here]}
	\end{center}

	The coefficient of $ D_{0} $ is positive and significant for the baseline specification. This result, however, is reversed once we allow for spillover effects by augmenting the baseline specification with measures of external knowledge. Since, as discussed in the next subsection, the spillover effects are particularly significant for the firms that do not engage in R\&D, it is likely that $ D^{0} $, which is a dummy for zero R\&D investment, is capturing the spillover effects in the baseline specification. This then suggests that firms that do not engage in R\&D are less productive.

	We also find that both incumbents and entrants experience larger productivity gains from additional capital accumulation than R\&D expenditure, suggesting that in Estonia `embodied technological change' through capital accumulation, most of which is imported, has been more effective in generating productivity growth \citep{castellani:2019}. Importantly, the average elasticity with respect to capital for R\&D performing firms was found to be higher than for those firms that did not engage in R\&D. For incumbents, the estimated elasticity with respect to capital for R\&D performing firms was found 1 to 3 percentage points higher than for those firms that did not engage in R\&D. As the test of equality of elasticity of productivity with respect to capital for firms with and without R\&D has been rejected (see Table \ref{tabres1}), these differences, though small, are significant. The fact that R\&D performing firms gain more from capital accumulation than non-R\&D firms suggests that R\&D helps firms build absorptive capacity, which  enables more efficient transfer of technology embodied in capital good \citep{griffith:2004,aghion:2015}.

	Comparing elasticity of productivity with respect to capital for the two countries in Table \ref{tabres3}, we find that, the elasticity is higher for the Estonian incumbents as compared to the German ones. Now, \citet{ortega:2014,ortega:2015} find that in traditional low-tech industries, which focus on process innovation, productivity gains turn out to be more related to capital accumulation rather than to R\&D expenditures. Supporting this argument, \citet{castellani:2019} point out that complex and radical product innovation generally relies on formal R\&D, while process innovation is much more related to embodied technological change achieved by investment in new machinery and equipment \citep[see also][]{parisi:2006}. In Estonia, (i) 47\% of the firm-year observations from low-tech and medium-low-tech manufacturing and 23\% are from less knowledge intensive services, and (ii) process innovation being more prevalent than product innovation (Table \ref{table:ds2}). This could potentially explain why productivity gains from capital accumulation is (a) higher than that from R\&D, and (b) higher for the Estonian incumbents as compared to the German ones.

	We find that the coefficient estimates of log(Employees) is negative for all firms. Now, in the value-added production function in (\ref{eq:5}), the coefficient of log(Employees) for R\&D performing firms is  $ \beta^{+}_{l} = \beta_{l}+ \beta_{k} + \beta_{r} - 1 $ and  $ \beta^{0+}_{l} = \beta^{0}_{l}+ \beta^{0}_{k}  - 1 $ for non-R\&D firms. The negative estimates of $ \beta^{+}_{l} $ and $ \beta^{0+}_{l}$, then suggest that the cumulative share of some of the inputs is less than 1. In other words, it suggests that the returns to scale is decreasing. As far as age is concerned, the estimates suggest that,  within the two groups of firms, age does not play a significant role as determinant of productivity. As a positive coefficient of age is associated with learning effects which causes improvements in productivity with time \citep{lubczyk:2020},\footnote{A negative coefficient of firm age, at least  for incumbents, indicates that these learning effects associated with firm age become much smaller and phase out in later stages of firm life.} the \citetalias{ackerberg:2015} estimates suggest that there are no learning effects.  Finally, we also find a robust evidence that firms located in northern Estonia, which is the economic hub of the nation, are more productive; this suggests that firms in northern Estonia benefit from a positive agglomeration effect.

	%Interestingly, in our data, knowledge-intensive services constitute about 15\% of the firm-year observations, which is high compared to the EU average \citep[see][p. 292, which reports a similar finding]{eu:2013}. However, as we found in our data, fewer firms in knowledge-intensive services introduced innovations, product and/or process, compared to firms in high-tech and medium-high-tech manufacturing.   

	\subsection{Spillover Effects}\label{sec52}

	Knowledge spillover effects, as we know, are highly relevant for both firms and policy makers, as they point to situations where the benefits of accumulating knowledge in one firm result in performance and productivity gains in a larger agglomeration or group of firms, further resulting in a sub-optimal low level of R\&D from a social point of view. Therefore, an important question related to the productivity effects of R\&D activities is the degree to which a firm's own R\&D efforts impact other firms' performances. Research on knowledge spillovers at the micro-level finds that the different mechanisms through which spillovers occur are indeed localised to a large extent \citep[see][]{audretsch:1998, storper:2004, ponds:2010}. Besides the importance of local labour markets and spin-off dynamics, studies have emphasized the role of networking between individuals and between organizations as a mechanism for knowledge spillovers that takes place at the regional level.  For each firm, then, we define external knowledge capital as the accumulation of knowledge that is generated by other firms that are geographically close to the focal firm.  
	
	Since knowledge spillovers can take place both within and between industries, we further differentiate the group of other firms into those firms that are within and those that are outside of the focal firm's own industry, and accordingly construct measures of inter- and intra-industry knowledge capital (see section \ref{sec4} for the definition of the two). These measures of external knowledge capital are then added to the production functions in equations (\ref{eq:5}). To test if there is differential impact of knowledge spillover for R\&D performing firms and non-R\&D firms, we interact the measures with two binary variables: $ D^{0} $ and $ 1- D^{0}  $. 
	
	Now, in our data, we find that the average R\&D expenditure of firms is higher in counties/regions where there is a higher concentration (number of firms/geographical area of the county) of firms. Moreover, we find that there is a positive correlation between R\&D expenses and the average R\&D expenses  of other firms situated in the same region. It, therefore, seems that location is not a random but a deliberate choice made by the firms.  In other words, the measures of external knowledge capital, through the endogenous choice of location, and firm's total factor productivity, $ \omega_{jt} $, are correlated. The endogeneity of measures of external knowledge capital is accounted for by treating measures of external knowledge capital and the dummy for north Estonia as state variables when using control function methods.               
	
	Table \ref{tabres1}, columns 2a and 2b, illustrates the estimated results when measures of external knowledge capital are added to the productivity functions. The estimates suggest that it is mainly the incumbents who benefit from external knowledge produced outside the firms' boundaries. However, among the incumbents, the R\&D performing incumbents seem to gain from only within industry external knowledge, whereas the incumbents who do not engage in R\&D activities gain from external knowledge both within and without their own industry.\footnote{Instead of the measure based on geographical distance, future research should consider estimating productivity response to measures of inter-industry external knowledge, which are  based on `technological distance' as in \citet{bloom:2013} \citep[see also][on measuering spillovers]{hall:2010}.} The estimates also suggest it is mainly the incumbents who benefit from knowledge capital produced outside their own firm. In other words, these results indicate that mature firms are on average better at incorporating diverse knowledge that stems from sources both within and outside their industry. The results also show that for the incumbents, the social returns to external knowledge capital are comparable to, if not higher than, the private returns to R\&D.

	We find that non-R\&D firms among the incumbents benefit more from external knowledge than innovating firms. Given that there exists complementarity between absorptive capacity and external knowledge and that at sufficiently low levels of absorptive capacity further increases in external knowledge may not increase marginal private incentives to build absorptive capacity to benefit from it \citep[see][]{aghion:2015}, the results indicate that incumbents that do not engage in R\&D activities, nonetheless do invest in building absorptivity capacity to benefit from external knowledge even though they do not invest in STI based innovation. Second, since by definition non-R\&D firms do not invest in `frontier innovation,' the results suggest that such incumbents are most likely benefiting from `technological adaptation or imitation,' which is aided by geographical proximity to R\&D intensive firms operating in the same industry.

	\subsection{Productivity Implication of Technological Innovation for Entrants and Incumbents} \label{sec53}
	As mentioned earlier, about 30\% of the firm-years have positive R\&D expenses, but about 75\% of the  firm-years in the sample have innovated to introduce at least one new product in the market or a new process. Also, about 7\% of the firms that invested in R\&D did not innovate a new product or process. These suggest that more firms innovate through the DUI mode of innovation rather than STI and that having positive R\&D expenses is neither a necessary nor a sufficient condition for innovation.
	
	Now, it has been found that firms in the low- and medium-tech sectors are usually engaged in the DUI mode of innovation while drawing on advanced science and technology results available through the common knowledge bases \citep{robertson:2009,trott:2017}. Given that the majority of firms in our sample are in the low-tech and low-medium-tech manufacturing and less knowledge intensive services (see Table \ref{table:ds3}), and many of whom innovate without formally engaging in R\&D, we, by estimating the average treatment effect (ATE) of technological innovation, $ \Delta $, given in equation (\ref{eq:10}), assess how technological innovations affect total factor productivity (TFP) and labour productivity consequently. As discussed in Appendix \ref{ap_a}, we first estimate the residual, $\omega_{jt} + \epsilon_{jt} = g(\omega_{j,t-1}, \mathcal{P}^{D}_{jt}, \mathcal{P}^{C}_{jt}) + \xi_{jt} + \epsilon_{jt}$, in equation (\ref{eq:5}) as a measure of TFP, and then use a matching method to estimate $ \Delta $.  In Table \ref{tabres2} we report the estimated ATEs.

	\begin{center}
		\textbf{[Table \ref{table:ds4} about here]}
	\end{center}

	\begin{center}
		\textbf{[Table \ref{tabres2} about here]}
	\end{center}

	To assess if there is complementarity between product and process innovation and to understand how (a) \textit{only} product innovation (b) \textit{only} process innovation and (c) \textit{both} product \textit{and} process innovation contribute to $ \Delta $, we also estimate the ATE of only product innovation ($ \Delta_{10} $), the ATE of only process innovation ($ \Delta_{01} $), and the ATE of product \textit{and} process innovation ($ \Delta_{11} $) (see equation (\ref{eq:12})). The control variables on the basis of which we match innovators with the non-innovators for estimating the ATEs in (\ref{eq:10}) and (\ref{eq:12}) are stated in the following. For estimating  $ \Delta $ in (\ref{eq:10}), we match `exactly' on year dummies (2006 to 2018) and 2-digit industry dummies, and match to `nearest neighbourhood' on the lags of the following variables: (1) residual, $\omega_{jt} + \epsilon_{jt}$, as a proxy for TFP, (2) estimate of $ \omega_{jt} $ (see footnote \ref{fn_ap} in Appendix \ref{ap_a}), (3) product innovation, (4) process innovation (5) technological innovation (6) log of R\&D intensity  (7) dummy for positive R\&D, (8) labour productivity, (9)  log of number of employees, (10) log of capital stock, (11) log of material cost, (12) log of intra-industry external knowledge, (13) log of inter-industry external knowledge, and (14) age.   However, because there were much fewer firms who only innovated products, only innovated processes, or both, we could not match exactly on industry dummies while estimating $ \Delta_{10} $, $ \Delta_{01} $, and $ \Delta_{11} $. We, therefore, matched exactly on year and to nearest neighbourhood on the rest of the variables while estimating $ \Delta_{10} $, $ \Delta_{01} $, and $ \Delta_{11} $.             
	
	As we can see in Table \ref{tabres2}, for productivity estimates from both the specifications, the ATE, $ \Delta $, of technological innovation on productivity is larger for the entrants as compared to the incumbents'. We find that the ATE ranges from 0.19 to 0.27 for the entrants, and same for the incumbents ranges from 0.1 to 0.13. These ATEs, however, are in logarithmic scale; at the mean, these difference in linear scale implies that an entrant who innovates is about 21\% to 30\% more productive than an entrant who does not. For the incumbents, it implies that, on average, the incumbent firm that innovates is about 10\% to 13\% more productive than the one which does not. 
	
	Given that a higher percentage of entrants (78\%) compared to incumbents (72\%) reported innovation, our results suggests that not only are entrants more innovative, but that entrants are abler than incumbents in translating their innovations into productivity gains. To understand how the two group of producers benefit from technological innovations, we look at a more disaggregated picture by estimating the ATE of only product innovation  ($ \Delta_{10} $), the ATE of only process innovation  ($ \Delta_{01} $), and the ATE of product \textit{and} process innovation ($ \Delta_{11} $). In Table \ref{tabres2}, we can see that $ \Delta_{10} $ is significantly positive for incumbents, implying that incumbents who innovated new products only realized productivity gains. And with $ \Delta_{01} $ significantly positive for entrants, it implies that entrants who implemented innovative processes only realized productivity gains. However, those firms that have innovated both product \textit{and} process realized the highest gain. The productivity of entrants who innovated both product \textit{and} process is about 35\% higher than an average entrant's, and such incumbents' productivity is  about 13\% higher than the average incumbent's.    
	
	Since for the incumbents, the ATE of process innovation only did not turn out significant, we can assume that for the incumbents $ \Delta_{01} =0$; similarly we can assume that $ \Delta_{10} $ is 0 for the entrants. We, therefore, find that for both the group of producers: $ \Delta_{11} > \Delta_{10} + \Delta_{01} $. In other words, we find the existence of complementarity between product and process innovations. Now, as can be seen in Table \ref{table:ds4}, a higher percentage of innovating entrants (50\%) and incumbents (40\%) innovate both products \textit{and} processes. That is, both the groups of producers, especially the entrants, prefer to invest in both the activities at the same time to maximize their productivity.      
	
	Complementarity between product and process innovations arises when the returns to implementing a product innovation are highest when the firm also implements a process innovation. This occurs because product innovations, which shift out the demand curve, induces the firm to produce a higher quantity. However, the returns from lowering the unit cost through process innovations are highest precisely when quantity produced is highest, and thus firms tend to implement product and process innovations together. 
	
	\citet{athey:1995} emphasize that flexibility in product designs and processes, which lowers the adjustment costs of implementing product and process innovations, enhances the complementarity between product and process innovations.  Since (a) entrants are found to have gained from implementing process innovations and (b) complementarity between product and process innovations is higher for the entrants, these results indicate that entrants do leverage their flexible and non-hierarchical structures --- due to which they lack the structural inertia for reorganisation and which allow for faster decision-making processes --- for implementing the complementary process innovations, and consequently realize a higher complementarity.    
	
	%That the ATE of only product innovation, $ \Delta_{10} $, is not significant for the entrants, does not imply that entrants do not gain from product innovations. Entrants gain from product innovation when they also innovate the complementary processes and implement them; for such entrants, the demand of their novel products may well be high. However, we cannot  separate the gains made due to increase in the firm's demand and the gains realized due to the lowering of costs.     
	
	% Given that a higher percentage of entrants (56\%) compared to incumbents (45\%) reported product innovation, it suggests that entrants are more likely to make productivity gains by widening their spectrum of final goods and/or by establishing niche markets. Furthermore, as it has been argued, because entrants are more inclined to exploit new ideas and invest in radical innovation, product innovations of the entrants yield greater rewards than that of the incumbents.

	Before proceeding further, we would like to point out that since a smaller percentage of firms --- incumbents (31\%) and entrants (29\%) --- reported engaging in formal R\&D, this result suggests that a lot of the productivity gains is being made through DUI mode of innovation. This assertion is also supported by our data: `DUI Exclusive mode of cooperation' or cooperation with other enterprises for innovative activities excluding R\&D was higher (53\%) for innovating firms that reported zero R\&D, whereas `STI Exclusive mode of cooperation' was higher (73\%) for firms with positive R\&D. While our data indicates that many innovations are likely a result of DUI mode of innovation, we are unable to identify which innovations are a result of DUI mode of innovation and which are STI based. We are thus unable to estimate the efficacy of STI based innovations vis-\`{a}-vis  DUI based innovations, or for that matter the efficacy of the combination of two, for the two group of producers. We leave the exercise of assessing the efficacy of STI and DUI based innovations for future research.
	%\footnote{This is because our panel data is unbalanced and we are unable to tell with certainty if the reported innovation of a firm with zero R\&D is not due to formal R\&D undertaken in the unobserved past. Besides, even for firms that engage in STI mode of innovation, some innovation could be due to the DUI mode and some could be due a combination of both. \citet{jensen:2007} have shown that firms that combine the two modes are likely to be more innovative than those relying primarily on one mode or the other. We leave the exercise of assessing the efficacy of STI based innovations vis-\`{a}-vis  DUI based innovations for future research.}

	Finally, we compare the estimated TFP for the entrants and incumbents.  As can be seen in Table \ref{tabres2}, on average, the TFP of the entrants is higher than that of the incumbents. This is in contrast to labour productivity, which on average is found to be higher among the incumbents than the entrants (see Table \ref{table:ds2}). Since labour productivity could change due to (a) changes in technology, which improves the efficiency with which the factors operate, or (b) due to changes in factor accumulation, it, therefore, seems that a higher labour productivity of the incumbents is because of a higher capital-labour ratio than due to more efficient technology. TFP, on the other hand, measures firm ability or efficiency, such as managerial talent, quality of inputs, innovation, etc., that are not accounted for by the observed inputs \citep[][]{chad:2011}. That surviving young firms have above average productivity and that they grow faster than their mature counterparts, has been documented elsewhere \citep{foster:2008,haltiwanger:2013}.

	Also, in Table \ref{tabres2} we can see that entrants' TFP have higher standard deviation as compared to the TFP for the incumbents. As discussed in \citet{haltiwanger:2013}, young firms exhibit an `up or out' dynamic --- they either grow fast on average or they exit. These `up or out' dynamics imply that exiting young firms have very low productivity while surviving young firms --- where  survival, more often than not, is because of successful innovation ---  have above average productivity.  Related to the `up or out' dynamics, \citet{foster:2018} explain that entry is associated `experimentation,'  which is critical for innovation and which results high degree of within-industry productivity dispersion. Following this experimentation phase is the `shakeout' period, in which entrepreneurs that successfully innovate and/or adopt grow while unsuccessful entrepreneurs contract and exit yielding productivity growth. \citet{foster:2018} discuss a variety of mechanisms which can help understand how `experimentation' can generate heterogeneity in the factors that cause dispersion.

	\section{Concluding Remarks}
	
	A large proportion of productivity growth, a key driver of economic growth, is due to reallocation emanating from the entry and exit of firms. Since this reallocation is induced by the selection effect due to the heterogeneous productivity impacts of R\&D investment and innovation undertaken by incumbents and entrants, in this paper we study the productivity implications of (a) own R\&D, (b) external knowledge, and (c) and technological innovations for incumbents and entrants. The third exercise is motivated by the fact that, in our sample, while only 30\% of firm-year report investment in R\&D, 74\% report technological --- product and process --- innovation. Studying the implications of R\&D alone would have provided a partial picture of the innovation activities of firms in Estonia and other CEE catching-up economies.  We, therefore, develop an empirical strategy to simultaneously estimate (1) the productivity elasticities of R\&D and external knowledge and (2) the impact of technological innovations on productivity. Also, in studying the productivity implication of technological innovation, we pay particular attention to complementarity between product and process innovation for the two groups of producers. Besides, since physical capital stock embodies technical change and innovation, we also study the productivity impact of capital accumulation.

	Several interesting conclusions emerge from our analysis. To start with, while both entrants and incumbents gain significantly from investing in R\&D, we do not find any difference in the average productivity elasticity with respect to own R\&D for the two group of producers. This is contrary to the findings in \citet{acemoglu:2018} and \citet{lubczyk:2020} for the technologically advanced economies, where entrants have been found to be disproportionately abler in translating their R\&D into productivity gains. Instead, we find that the impact of innovation output --- many of which are a result of DUI based innovation --- are similar to the results in the aforementioned papers: entrants who innovate are 21\% to 30\% more productive than entrants who do not; for the incumbents, the corresponding figures are 10\% to 13\%. Furthermore, as we explored the sources of the productivity gain, we found that the gain is largely due to complementarities between product and process innovation, with entrants benefiting more from the  complementarities than incumbents.

	Estonian incumbents, on the other hand, are no less abler than German incumbents in translating R\&D or STI based innovative activities into productivity gains. This is despite the adverse sectoral composition of the  Estonian economy, which has a low share of sectors that are R\&D intensive and where returns to R\&D are highest. Moreover, because the bulk of STI based innovative activities in Estonia is carried out by the incumbents, they are the primary carriers of STI based innovative activities.

	We find that productivity elasticity with respect to capital is higher than with respect to R\&D, suggesting that in catching-up economies such as in Estonia, embodied technological change through capital accumulation has been an important means, and likely more effective than R\&D, for generating productivity growth. However, since productivity elasticity with respect to capital is higher for firms that report engaging in formal R\&D, it suggests that STI based innovative activities do help build absorptive capacity.     
	
	Finally, in results related to spillover effects, we find it is mostly the incumbents who benefit from intra- and inter-industry knowledge that is generated by other firms --- primarily by the incumbents --- situated in close proximity. However, the spillover effects are higher for firms that do not reportedly engage in formal R\&D, suggesting that these firms, too, work on building absorptive capacity.
	
    There are certain limitations in our paper, which are primarily due to lack of data. First, as Community Innovation Surveys (CIS) consider firms with at least 10 employees, the entrants included in these data may not be representative of the population of newly born firms, which are unlikely to exceed this threshold in the first years of their existence. Relying on CIS data could therefore limit the number of newly established firms. Second, we consider only regional spillover effects; future studies could consider measures of external knowledge that are based on technological distance between firms and/or forward and backward linkages between industries.

	Our results have important policy implications. The finding that technological innovations benefit entrants disproportionately more than the incumbents points to the significant impact innovative entry can have on economic dynamics. These results yet again emphasize policies --- in particular those that facilitate DUI mode of innovation ---  for promoting innovative entrepreneurship. However, our results also strongly suggest than in Estonia --- a catching-up economy ---  incumbents are the primary carriers of STI based innovative activities, which are instrumental in building absorptive capacity and help move economies closer to the technological frontiers. 
	
	Our results then suggest that the policies advocated in \citet{acemoglu:2018} for technologically advanced economies may not be applicable for a catching-up economy like Estonia. \citet{acemoglu:2018} advocate that a large tax on the continued operation of incumbents combined with a modest R\&D subsidy to incumbent would be both growth and welfare increasing. However, this policy only leverages the selection effect. As pointed out by \citet[p. 539]{aghion:2015}, the recommendations suggested by \citet{acemoglu:2018} do not take into consideration the notion of absorptive capacity which could have ``novel implications, as it creates another trade-off between incumbents (with a large stock of R\&D and a high absorptive capacity) and new firms." Policy makers must, therefore, duly consider the important role incumbents, as primary carriers of STI based innovative activities, play in catching-up economies. As argued in \citet{aghion:2015}, because the social returns to absorptive capacity are not fully internalised by the private market, private market forces may fail to ensure the catching-up process; policies and institution that foster large incumbents during the catching-up periods have historical antecedents \citep[see][]{aghion:2010}.

	\bibliography{Citation}{}
\bibliographystyle{ecca}
	\clearpage
	
	\appendix
	
	\section*{Appendices}
	\addcontentsline{toc}{section}{Appendices}
	\renewcommand{\thesection}{\Alph{section}}
	\renewcommand{\thesubsection}{\Alph{subsection}}

	\section{Identification of Average Treatment Effect of Innovation and Testing for Complementarity Between Product and  Innovation}\label{ap_a}

	Substituting the specification for $ \omega_{jt} $ given in equation (\ref{eq:9}) in the log-linear production function in equation (\ref{eq:5}), we obtain 
	\begin{align}\nonumber
	y_{jt} = &  (1- D_{jt}^{0})(\beta+ \beta^{+}_{l}l_{jt} + \beta_{k}k_{jt}  + \beta_{r}r_{jt} +\beta_{e}e_{jt})   + D_{jt}^{0}(\beta^{0} + \beta^{0+}_{l}l_{jt} + \beta^{0}_{k}k_{jt} +\beta^{0}_{e}e_{jt} ) \\ \label{eq:a1} 
	& + g(\omega_{j,t-1}, \mathcal{P}^{D}_{jt}, \mathcal{P}^{C}_{jt}) + \xi_{jt} + \epsilon_{jt}.  
	\end{align}  
	The coefficients, $\beta$'s, and the function, $ g(\omega_{j,t-1}, \mathcal{P}^{D}_{jt}, \mathcal{P}^{C}_{jt}) $, as discussed in section 4, is estimated using the control function methodology by \citet{ackerberg:2015} (ACF). For notational convenience from here on, we drop the firm subscript, $ j $.            
	
	In this appendix, we discuss how we (i) estimate the average treatment effect (ATE) of the various kinds of innovation ($\Delta$ in Eq. (\ref{eq:10}) and $ \Delta_{10} $, $ \Delta_{01} $, $ \Delta_{11} $ in (\ref{eq:12}) ) and (ii) assess for complementarity between product and process innovation. 
	
	Now, as discussed in the main text, the function, $g(\omega_{t-1}, \mathcal{P}^{D}_{t}, \mathcal{P}_{t}^{C}) $, in equation (\ref{eq:a1}) is estimated non-parametrically by approximating it by a polynomial function of the arguments of $g(\omega_{t-1}, \mathcal{P}^{D}_{t}, \mathcal{P}_{t}^{C}) $. So, if, for example, $ \omega_{t-1} $ is calculated as in footnote \ref{fn_ap}, one could employ the estimated coefficients of the polynomial to obtain estimates of $ g_{1\vee1}(\omega_{t-1}, .) $ and $g_{0\wedge 0}(\omega_{t-1}, .)$, where the two are defined in equation (\ref{eq:11}), for every $ \omega_{t-1} $, and then by averaging over $ \omega_{t-1} $ one can obtain, for example, $ \Delta $, the average treatment effect of technological innovation: 
	 \begin{align}\label{eq:a0}\Delta = \E[g_{1\vee1}(\omega_{t-1}, .)] -  \E[g_{0\wedge 0}(\omega_{t-1}, .)].
	 \end{align}
For inference, we could either (a) first obtain bootstrapped standard errors of the coefficients of the polynomial function, and then use the delta method to obtain the standard error of $\Delta$, or (b) directly bootstrap $\Delta$. Both procedures are, however, computationally intensive.  Rather than following these laborious procedures of computing the ATEs and their standard errors, we, as discussed below, develop a simple procedure to estimate the ATEs.

In the following we show that we can still obtain $\Delta $ in (\ref{eq:a0}) without using the estimates of the function, $g(\omega_{t-1}, .)$. The alternative procedure entails first estimating the residuals \begin{align}\nonumber
r_{t} &\equiv y_{t} - \big( (1- D_{t}^{0})(\beta+ \beta^{+}_{l}l_{t} + \beta_{k}k_{t}  + \beta_{r}r_{t} +\beta_{e}e_{t})   + D_{t}^{0}(\beta^{0} + \beta^{0+}_{l}l_{t} + \beta^{0}_{k}k_{t} +\beta^{0}_{e}e_{t} ) \big)\\\nonumber
	&= g(\omega_{t-1}, \mathcal{P}^{D}_{t} , \mathcal{P}_{t}^{C}) + \xi_{t} + \epsilon_{t} 
	\end{align} in equation (\ref{eq:a1}) and then using the matching methods for estimating the ATE, $ \Delta $.  The alternative procedure works because these residuals, which are a proxy for the total factor productivity (TFP), contain information on $ g(\omega_{t-1}, .) $.  The residuals for firm-years that innovated are $ r_{t}(I_{t} = 1) = g_{1 \vee 1}(\omega_{t-1}, .) + \xi_{t} + \epsilon_{t} $, and for those that did not are $ r_{t}(I_{t} = 0) =  g_{0\wedge 0}(\omega_{t-1}, .) + \xi_{t} + \epsilon_{t} $.

	Having obtained the residuals, $ g(\omega_{t-1}, .) + \xi_{t} + \epsilon_{t} $, to estimate the impact of technological innovations on productivity, we can obtain the difference 
	\begin{align}\label{eq:a2}
	\E[g_{1 \vee 1}(\omega_{t-1}, .) + \xi_{t} + \epsilon_{t}] - \E[g_{0\wedge 0}(\omega_{t-1}, .) + \xi_{t} + \epsilon_{t}],
	\end{align}    
	where the expectation is taken over $ (\omega_{t-1}, \xi_{t}, \epsilon_{t}) $.  
	%That is,  \begin{align}\nonumber& g_{1 \vee 1}(.) \equiv g(\omega_{t-1}, \mathcal{P}^{D}_{t} , \mathcal{P}_{t}^{C}) \text{ such that } \mathcal{P}^{D}_{t} =1  \textit{ and/or } \mathcal{P}^{C}_{t}=1 \text{ and }\\ \label{eq:a3}& g_{0\wedge 0}(.) \equiv g(\omega_{t-1}, \mathcal{P}^{D}_{t} , \mathcal{P}_{t}^{C}) \text{ such that } \mathcal{P}^{D}_{t} =0  \textit{ and } \mathcal{P}^{C}_{t}=0.  \end{align} 
	
	Note that we only know $ r_{t}(I_{t} = 1) $ for firm-years that have innovated and $ r_{t}(I_{t} = 0)$ for firms that did not innovate. Our problem is, therefore, same as the missing data problem in treatment effect literature: $ r_{t}(I_{t} = 0) $ is unobserved for firm-years that innovated, and $ r_{t}(I_{t} = 1) $ is unobserved for firm-years that did not innovate. The comparison, therefore, between $\E[g_{1 \vee 1}(\omega_{t-1}, .) + \xi_{t} + \epsilon_{t}]$ and $ \E[g_{0\wedge 0}(\omega_{t-1}, .) + \xi_{t} + \epsilon_{t}] $ for assessing the effect of innovation can only be possible if $ \omega_{t-1} $ for firms that have innovated (treatment group) are same as $\omega_{t-1}$ for firms that did not (control group). 
	
	To select firms in control group with $\omega_{t-1}$ similar to the firms in the treatment group, we resort to the matching method in \citet{abadie:2006}. The method rigorously selects control units with $X_{t-1}$ similar to the treated units. The control variables (or `observable confounders'), $ X_{t-1} $, on the basis of which we match firms in treatment group to the firms in the control group are variables that likely affect both the treatment (innovation) and the outcome (TFP). In $ X_{t-1} $ we include (i) $ \omega_{t-1} $,\footnote{\label{fn_ap}In \citetalias{ackerberg:2015}, firm's intermediate input demand is given by $ m_{t} = f_{t}(l_{t}, k_{t}, r_{t}, e_{t}, x_{t}, \omega_{t}) $, where $ x_{t} $ is the set of additional state variables, and the function, $ m_{t} = f_{t}(., \omega_{t}) $, is strictly monotonic in the unobserved TFP, $ \omega_{t} $. Given the monotonicity assumption, a proxy for $ \omega_{t} $ is obtain by inverting the intermediate input demand function:  $ \omega_{t} = f^{-1}_{t}(l_{t}, k_{t},  r_{t}, e_{t}, x_{t}, m_{t}) $. STATA's prodest command gives an estimate of $\omega_{t} $ as \begin{align}\nonumber
		\hat{\phi}(D^{0}_{t},l_{t}, k_{t}, r_{t}, e_{t}, x_{t}, m_{t})&-[(1- D_{t}^{0})(\hat{\beta}+ \hat{\beta}^{+}_{l}l_{t} + \hat{\beta}_{k}k_{t}  + \hat{\beta}_{r}r_{t} +\hat{\beta}_{e}e_{t})   + D_{t}^{0}(\hat{\beta}^{0} + \hat{\beta}^{0+}_{l}l_{jt} + \hat{\beta}^{0}_{k}k_{t} +\hat{\beta}^{0}_{e}e_{t} )], 
		\end{align}  where $ \phi(.) $, which is estimated in the first stage, is given by 
		\begin{align}\nonumber
		\phi(D^{0}_{t},l_{t}, k_{t}, r_{t}, e_{t}, x_{t}, m_{t}) =& [(1- D_{t}^{0})(\beta+ \beta^{+}_{l}l_{t} + \beta_{k}k_{t}  + \beta_{r}r_{t} +\beta_{e}e_{t})   + D_{t}^{0}(\beta^{0} + \beta^{0+}_{l}l_{t} + \beta^{0}_{k}k_{t} +\hat{\beta}^{0}_{e}e_{t} )]\\\nonumber&+f^{-1}_{t}(l_{t}, k_{t}, r_{t}, e_{t}, x_{t}, m_{t}).
		\end{align} Note that given $ \omega_{t} = f^{-1}_{t}(l_{t}, k_{t},  r_{t}, e_{t}, x_{t}, m_{t}) $, the next period productivity is  $ \omega_{t+1} = g(\omega_{t}, \mathcal{P}^{D}_{t}, \mathcal{P}^{C}_{t}) + \xi_{t}$, and it is $ g(\omega_{t}, \mathcal{P}^{D}_{t}, \mathcal{P}^{C}_{t}) $ that is our quantity of interest, not $ \omega_{t} = f^{-1}_{t}(l_{t}, k_{t},  r_{t}, e_{t}, x_{t}, m_{t}) $.} (ii) all the state variables and control variables from period $ t-1 $, and (iii) $ \mathcal{P}^{D}_{t-1} $ and $ \mathcal{P}^{C}_{t-1} $. The control variables, $ X_{t-1} $, are mean independent of the error terms, $ \xi_{t} + \epsilon_{t} $ (see also discussion following Eq. (\ref{eq:9})).

	The identification assumption is that conditional on $ X_{t-1} $, the treatment is independent of the `potential' outcome, $ r_{t}(I_{t}) $. In the other words, for the matched sample, the treatment (here innovation) can then be considered to be randomly assigned between the treatment and the control group.  Under this assumption, for the matched sample, we, therefore, have
	\begin{align}\nonumber
	\Delta(X_{t-1}) &= \E(r_{t}(I_{t} = 1) -r_{t}(I_{t} =0)| X_{t-1}  )\\\nonumber
	&= \E[g(\omega_{t-1}, .) + \xi_{t} + \epsilon_{t}|I_{t} = 1, X_{t-1}] - \E[g(\omega_{t-1}, .) + \xi_{t} + \epsilon_{t}|I_{t} = 0, X_{t-1}]  \\\nonumber
	&= \E[g_{1 \vee 1}(\omega_{t-1}, .) + \xi_{t} + \epsilon_{t}| X_{t-1}] - \E[g_{0\wedge 0}(\omega_{t-1}, .) + \xi_{t} + \epsilon_{t}| X_{t-1}] \\\label{eq:a3}
	&= \E[g_{1 \vee 1}(\omega_{t-1}, .)|  X_{t-1}  ]  - \E[g_{0\wedge 0}(\omega_{t-1}, .)| X_{t-1}], 
	\end{align}
 where the last equality follows also because $\xi_{t} +\epsilon_{t}$ is mean independent of $ X_{t-1} $.

	Under the overlap assumption, $ \pr(I_{t}=1|X_{t-1}) <1 $, \citep[see][]{abadie:2006} the difference on the right-hand side
	of (\ref{eq:a3}) is identified for almost all $ X_{t-1} $ in the support of $ X_{t-1} $. Therefore, the average effect of the treatment can be recovered by averaging $ \E[g_{1 \vee 1}(\omega_{t-1}, .)|  X_{t-1}  ]  - \E[g_{0\wedge 0}(\omega_{t-1}, .)| X_{t-1}] $ over the distribution of $ X_{t-1} $. That is, 
	\begin{align}\label{eq:a4}
	\E[\Delta(X_{t-1})]   = \E[g_{1 \vee 1}(\omega_{t-1}, .)]  - \E[g_{0\wedge 0}(\omega_{t-1}, .)] = \Delta, 
	\end{align}
	which is what we wanted to estimate in (\ref{eq:a0}).

	We now come to assessing of complementarity between product, $ \mathcal{P}^{D}_{t} =1 $, and process, $\mathcal{P}^{C}_{t}=1  $ innovation. The study of complementarities between activities can be traced back to the theory of supermodularity \citep{milgrom:1990}. According to this theory, the necessary condition for product and process innovation to be complementary is that the expectation, $\E[g(\omega_{t-1}, \mathcal{P}^{D}_{t}, \mathcal{P}^{C}_{t})]$, be supermodular in $\mathcal{P}^{D}_{t}$ and $ \mathcal{P}^{C}_{t}$: 
	\begin{align}\label{eq:a5}
	\E[g_{1 \wedge 1}(.)] -\E[g_{0 \wedge 1}(.)]\geq
	\E[g_{1\wedge 0}(.)] - \E[g_{0 \wedge 0}(.)],
	\end{align}
	where
	\begin{align}\nonumber
	& g_{1 \wedge 0}(.) \equiv g(\omega_{j,t-1}, \mathcal{P}^{D}_{jt} , \mathcal{P}_{jt}^{C}) \text{ such that } \mathcal{P}^{D}_{jt} =1  \textit{ and  } \mathcal{P}^{C}_{jt}=0, \\\nonumber 
	& g_{0 \wedge 1}(.) \equiv g(\omega_{j,t-1}, \mathcal{P}^{D}_{jt} , \mathcal{P}_{jt}^{C}) \text{ such that } \mathcal{P}^{D}_{jt} =0  \textit{ and } \mathcal{P}^{C}_{jt}=1,   \\ \nonumber
	& g_{1 \wedge 1}(.) \equiv g(\omega_{j,t-1}, \mathcal{P}^{D}_{jt} , \mathcal{P}_{jt}^{C}) \text{ such that } \mathcal{P}^{D}_{jt} =1  \textit{ and } \mathcal{P}^{C}_{jt}=1 \text{ and }  \\ \nonumber
	& g_{0 \wedge 0}(.) \equiv g(\omega_{j,t-1}, \mathcal{P}^{D}_{jt} , \mathcal{P}_{jt}^{C}) \text{ such that } \mathcal{P}^{D}_{jt} =0  \textit{ and } \mathcal{P}^{C}_{jt}=0. 
	\end{align} 
	The inequality in (\ref{eq:a5}) can be interpreted as follows: on average, implementing a product innovation when the firm also implements a process innovation has a higher incremental effect on productivity than implementing a product innovation in isolation.
	
	We can write the inequality in (\ref{eq:a5}) as 
	\begin{align} \nonumber
	&\E[g_{1 \wedge 1}(.)] - \E[g_{0 \wedge 0}(.)] &\geq& (\E[g_{0 \wedge 1}(.)] - \E[g_{0 \wedge 0}(.)]) &+& 
	(\E[g_{1\wedge 0}(.)] - \E[g_{0 \wedge 0}(.)]) \\\label{eq:a6}
	\Longleftrightarrow & \quad \qquad \Delta_{11} &\geq& \quad \qquad \Delta_{01} &+& \quad \qquad  \Delta_{10},
	\end{align}
	where $ \Delta_{10} $  is the ATE of only product innovation, $ \Delta_{01} $ is the ATE of only process innovation, and  $ \Delta_{11} $, the ATE of both product \textit{and} process innovation.  That is, if the productivity impact of product \textit{and} process innovation exceeds the sum of the impacts of only product innovation and only process innovation, then the product and process innovation can be said to be complementary. The ATEs, $ \Delta_{10} $, $ \Delta_{01} $, and  $ \Delta_{11} $, can be estimated using the matching method as in (\ref{eq:a4}).

\vspace{3cm}	
	
	% add 'Appendix' to the section heading
\newcommand{\appsection}[1]{\let\oldthesection\thesection
	\renewcommand{\thesection}{Section \oldthesection}
	\section{#1}\let\thesection\oldthesection}

	\renewcommand{\thetable}{\arabic{table}}
	% redefine the command that creates the equation no.
	\setcounter{table}{0}  % reset counter
	%\section*{APPENDIX}  % use *-form to suppress numbering

	%\section{Tables and Figures}

	\begin{table}[h!]
		
		\caption{\textbf{Test of Equality of Means between Entrants and Incumbents }}
		\centering
		\resizebox{\textwidth}{!}{	\begin{tabular}{l |c |c |c |c }
				\hline\hline
				& Entrants & Incumbents & Difference & $ \pr(|T|>|t|) $\\ \hline
				log(Value-Added Productivity )	&	10.2	&	10.33	&	-0.13	&	0.00	\\
				log(Gross Output Productivity)	&	11.16	&	11.27	&	-0.11	&	0.00	\\
				log(No. of Employees) 	&	3.5	&	3.75	&	-0.25	&	0.00	\\
				log(Capital) 	&	12.4	&	13.36	&	-0.96	&	0.00	\\
				log(Material Cost) 	&	13.57	&	14.01	&	-0.44	&	0.00	\\
				log(Investment in Fixed Assets)	&	11.17	&	11.53	&	-0.36	&	0.00	\\
				log(R\&D Expenditure) 	&	7.17	&	7.78	&	-0.61	&	0.000	\\
				Dummy for Positive R\&D Expenditure	&	0.29	&	0.31	&	-0.02	&	0.16	\\
				Dummy for Innovator	&	0.78	&	0.72	&	0.06	&	0.00	\\ 
				Dummy for Product Innovation	&	0.56	&	0.45	&	0.11	&	0.00	\\ 
				Dummy for Process Innovation	&	0.61	&	0.56	&	0.04	&	0.00	\\ 
				Dummy for North	& 0.59 & 0.54 & 0.05 & 0.00 \\
				log(Intra-Industry R\&D)	&	4.27	&	4.56	&	-0.28	&	0.02	\\
				log(Inter-Industry R\&D)	&	9.39	&	10.07	&	-0.68	&	0.00	\\
				\hline \hline
				\multicolumn{5}{p{15cm}}{Note: (a) The two sample t-test assumed unequal variances. (b) The data used for the tests is the estimation sample. } \\
				
		\end{tabular} }
		\label{table:ds2}
	\end{table}

	\begin{table}[h!]
		\caption{\textbf{Sectoral Distribution of Estonian Firms in the Estimation Sample}}
		\centering
		\begin{adjustbox}{max width=\textwidth}
			\begin{tabular}{l|c|c|c|c} 
				\hline \hline
				&	Percentage  	&	Positive R\&D	&	Technological 	& \multirow{2}{*}{$ \log\Big(\dfrac{\text{R\&D}}{\text{Employees}}\Big) $}		\\
				&	of Firms	&		&	Innovation 	&		\\ \hline
				&\multicolumn{4}{c}{Entrants} \\\hline
				High-Tech Manufacturing 	&	2.3	&	37.5	&	97	&	3.41	\\
				Medium High-Tech Manufacturing	&	10.14	&	35.5	&	86	&	2.54	\\
				Medium Low-Tech Manufacturing	&	22.01	&	22.9	&	75	&	1.57	\\
				Low-Tech Manufacturing	&	28.42	&	27.1	&	81	&	1.84	\\
				Knowledge Intensive Services 	&	19.93	&	43	&	82	&	3.47	\\
				Less Knowledge Intensive Services 	&	17.19	&	19.2	&	69	&	1.15	\\ \hline
				Total 	&	100	&	28.9	&	78	&	2.1	\\ \hline
				& \multicolumn{4}{c}{Incumbents} \\\hline
				High-Tech Manufacturing 	&	2.26	&	54.1	&	83	&	4.7	\\
				Medium High-Tech Manufacturing	&	11.02	&	46	&	82	&	3.78	\\
				Medium Low-Tech Manufacturing	&	15.13	&	27.1	&	71	&	2	\\
				Low-Tech Manufacturing	&	31.62	&	28.2	&	77	&	1.94	\\
				Knowledge Intensive Services 	&	15.59	&	41	&	72	&	3.52	\\
				Less Knowledge Intensive Services 	&	24.39	&	21.5	&	64	&	1.68	\\\hline
				Total 	&	100	&	30.7	&	72	&	2.4	\\
				
				\hline \hline 
				\multicolumn{5}{l}{No. of Entrants: 1500, No. of Incumbents: 6086.} \\
				\multicolumn{5}{p{15cm}}{Note: (a) All figures excepts for log(R\&D/Employees) are in percentages. (b)  Technological classification borrowed from Eurostat is based on NACE Revision 2 2-digit level. } 
				
			\end{tabular}
		\end{adjustbox} \label{table:ds3}
	\end{table}

	\begin{table}[h!]	
		
		\caption{\textbf{Productivity Effects of R\&D and Capital Accumulation for Entrants and Incumbents}}															
		\centering																	
		\begin{adjustbox}{max width=\textwidth}
			\begin{tabular}{l|r@{}l | r@{}l| r@{}l| r@{}l}																	
				\hline\hline															
				&	 \multicolumn{4}{c|}{Specification: Baseline} 					&	  \multicolumn{4}{c}{Specification: Baseline + } 									\\\hline
				&	 \multicolumn{4}{c|}{} 					&	  \multicolumn{4}{c}{ External Knowledge} 									\\\hline
				&\multicolumn{2}{c|}{Incumbent}& \multicolumn{2}{c|}{Entrant}& \multicolumn{2}{c|}{Incumbent}& \multicolumn{2}{c}{Entrant} \\\hline  
				&\multicolumn{2}{c|}{(1a)}& \multicolumn{2}{c|}{ (1b)}& \multicolumn{2}{c|}{ (2a)}& \multicolumn{2}{c}{(2b)} \\\hline
				log(Employees)$ \times(1-D^{0}) $ 	&	-0.16	&	\onepc	&	-0.035	&		&	-0.234	&	\onepc	&	-0.188	&	\onepc		\\
				&	(0.001)	&		&	(0.011)	&			&	(0.001)	&		&	(0.013)	&	\\
				log(Employees)$ \times D^{0} $ 	&	-0.147	&	\onepc	&	-0.109	&	\onepc				&	-0.141	&	\onepc	&	-0.101	&	\onepc			\\
				&	(0.001)	&		&	(0.03)	&			&	(0.001)	&		&	(0.007)	&					\\
				log(Capital/Employees) $ \times(1-D^{0}) $  	&	0.304	&	\onepc	&	0.194	&	\onepc		&	0.300	&	\onepc	&	0.186	&	\onepc		\\
				&	(0.002)	&		&	(0.023)	&				&	(0.001)	&		&	(0.009)	&			\\
				log(Capital/Employees) $ \times D^{0} $ 	&	0.293	&	\onepc	&	0.198	&	\onepc		&	0.289	&	\onepc	&	0.174	&	\onepc		\\
				&	(0.002)	&		&	(0.016)	&				&	(0.002)	&		&	(0.007)	&		\\
				log(R\&D/Employees)	&	0.046	&	\onepc	&	0.064	&	\onepc	&	0.081	&	\onepc	&	0.079	&	\onepc		\\
				&	(0.005)	&		&	(0.02)	&				&	(0.002)	&		&	(0.009)	&		\\
				$ D^{0}  $:  Dummy Zero R\&D	&	0.047	&	\onepc	&	0.067	&	\onepc	&	-0.308	&	\onepc	&	-0.276	&	\onepc		\\
				&	(0.001)	&		&	(0.009)	&				&	(0.001)	&		&	(0.013)	&				\\
				log(Intra-industry R\&D)$ \times(1-D^{0}) $ 	&		&		&		&				&	0.022	&	\onepc		&	0.019  & \onepc	\\
				&		    &    &			&		&	(0.003)	&		&			(0.001)	&			\\
				log(Intra-industry R\&D)$ \times D^{0} $ 				&		&		&		&		&	0.025	&	\onepc	&	0.01	&			\\
				&		&		&		&			&	(0.002)	&				&	(0.001)	&			\\
				log(Inter-industry R\&D)$ \times(1-D^{0}) $ 	&		&		&		&		&	-0.009	&		&	-0.04	&			\\
				&		&		&		&			&	(0.007)	&		&	(0.009)	&		\\
				log(Inter-industry R\&D) $ \times D^{0} $ 	&		&		&		&				&	0.035	&	\onepc	&	0.015	&	\fivepc		\\
				&		&			&		&		&	(0.001)	&				&	(0.001)	&				\\
				Age 	&	0.000	&		&	0.017	&		&	0.003	&		&	-0.005	&		\\
				&	(0.001)	&		&	(0.019)	&				&	(0.002)	&		&	(0.008)	&			\\
				North Estonia 	&	0.326	&	\onepc	&	0.268	&	\onepc		&	0.192	&	\onepc	&	0.314	&	\onepc			\\
				&	(0.002)	&		&	(0.009)	&				&	(0.000)	&		&	(0.009)	&			\\ \hline
				\multicolumn{9}{p{15cm}}{Test of Equality of Elasticity of Productivity with respect to Capital for firms with and without R\&D.}\\\hline
				Test Statistic: $ C $	&	6.92 & \onepc			&	\multicolumn{2}{c|}{0.06}		&	6.03	&\fivepc		&	\multicolumn{2}{c}{0.78}				\\		
				$ \pr ( \chi^{2}(1)>C) $ &	\multicolumn{2}{c|}{0.008}			&	\multicolumn{2}{c|}{0.799}						&	\multicolumn{2}{c|}{0.014}			&	\multicolumn{2}{c}{0.376}			\\\hline			
				No. of Observations	&	\multicolumn{2}{c|}{6086}			&	\multicolumn{2}{c|}{1500}						&	\multicolumn{2}{c|}{5877}			&	\multicolumn{2}{c}{1464}					
				\\
				No. of Firms	&	\multicolumn{2}{c|}{2443}			&	\multicolumn{2}{c|}{1072}							&	\multicolumn{2}{c|}{2394}			&	\multicolumn{2}{c}{1052}			 \\
				\hline 
				\hline
				\multicolumn{9}{p{15cm}}{Note: Dependent Variable: log(Value-Added/Employees). Every specification includes Time and Industry Dummies.}	\\															
				\legend  																	
			\end{tabular} 
		\end{adjustbox}															
		\label{tabres1}																	
	\end{table}

	\begin{sidewaystable}[h!]																	
		\caption{\textbf{Comparison of Elasticity of Productivity with respect to Own R\&D and Capital with those obtained for Germany  }}					
		\centering 														
		\resizebox{\textwidth}{!}{	\begin{tabular}{l|r@{}l | r@{}l| r@{}l| r@{}l||r@{}l | r@{}l | r@{}l |r@{}l}																	
				\hline\hline															
				&	 \multicolumn{8}{c||}{Specification: Baseline} 					&	  \multicolumn{8}{c}{Specification: Baseline + External Knowledge} 									\\\hline
				&	 \multicolumn{4}{c|}{Method: \citetalias{ackerberg:2015}}	&	\multicolumn{4}{c||}{Method: \citetalias{olley:1996}}	&	 \multicolumn{4}{c|}{Method: \citetalias{ackerberg:2015}}	&	\multicolumn{4}{c}{Method: \citetalias{olley:1996}}	\\\hline 
				&\multicolumn{2}{c|}{Incumbent}& \multicolumn{2}{c|}{Entrant}& \multicolumn{2}{c|}{Incumbent}& \multicolumn{2}{c||}{Entrant}& \multicolumn{2}{c|}{Incumbent}& \multicolumn{2}{c|}{Entrant}&
				\multicolumn{2}{c|}{Incumbent}& \multicolumn{2}{c}{Entrant}\\\hline  
				&\multicolumn{2}{c|}{(1a)}& \multicolumn{2}{c|}{ (1b)}& \multicolumn{2}{c|}{(1c)}& \multicolumn{2}{c||}{(1d)}& \multicolumn{2}{c|}{ (2a)}& \multicolumn{2}{c|}{(2b)}&
				\multicolumn{2}{c|}{(2c)}& \multicolumn{2}{c}{(2d)}\\\hline
				\multicolumn{17}{c}{Estonia}	\\\hline
				log(Capital/Employees) $ \times(1-D^{0}) $  	&	0.304	&	\onepc	&	0.194	&	\onepc	&	0.179	&	\onepc	&	0.164	&	\onepc	&	0.300	&	\onepc	&	0.186	&	\onepc	&	0.193	&	\onepc	&	0.128	&	\onepc	\\
				&	(0.002)	&		&	(0.023)	&		&	(0.017)	&		&	(0.057)	&		&	(0.001)	&		&	(0.009)	&		&	(0.002)	&		&	(0.033)	&		\\
				log(Capital/Employees) $ \times D^{0} $ 	&	0.293	&	\onepc	&	0.198	&	\onepc	&	0.148	&	\onepc	&	0.116	&	\fivepc	&	0.289	&	\onepc	&	0.174	&	\onepc	&	0.176	&	\onepc	&	0.074	&	\onepc	\\ 
				&	(0.002)	&		&	(0.016)	&		&	(0.017)	&		&	(0.051)	&		&	(0.002)	&		&	(0.007)	&		&	(0.002)	&		&	(0.027)	&		\\ \hline
				log(R\&D/Employees)	&	0.046	&	\onepc	&	0.064	&	\onepc	&	0.026	&	\fivepc	&	0.032	&	\fivepc	&	0.081	&	\onepc	&	0.079	&	\onepc	&	0.062	&	\onepc	&	0.037	&	\fivepc	\\
				&	(0.005)	&		&	(0.02)	&		&	(0.012)	&		&	(0.015)	&		&	(0.002)	&		&	(0.009)	&		&	(0.002)	&		&	(0.009)	&		\\ \hline \hline
				\multicolumn{17}{c}{Germany} \\\hline
				log(Capital/Employees)		& 0.127 & \onepc & 0.281 & \onepc & 0.076 & \onepc &   0.147 & \onepc 
				& 0.133 & \onepc & 0.255 & \onepc &  0.021 & & 0.148 & \onepc \\
				& (0.000) &      & (0.001)&       & (0.026) & & (0.029) & 
				& (0.002) &      & (0.003) &      & (0.026) & & (0.007) & \\\hline 
				log(R\&D/Employees)	& 	0.057 & \onepc & 0.125 & \onepc &  0.024 & \onepc &0.029 & \onepc 
				&   0.050 & \onepc & 0.120 & \onepc & 0.022 & \onepc & 0.026 & \onepc  \\
				&	(0.002) & & (0.002) & & (0.004) & & (0.010) & 
				&   (0.002) & & (0.003) & & (0.002) & & (0.009) & \\
				\hline \hline
				\multicolumn{17}{p{23cm}}{Note: \citet{lubczyk:2020} estimated the elasticities using two different control function methods: one, the method by \citetalias{ackerberg:2015}, and the other by \citet{olley:1996} (OP). For the purpose of comparison, we also estimate using both the methods. The dependent variable is log(Value-Added/Employees) when employing \citetalias{ackerberg:2015} and log(Revenue/Employees) when employing \citetalias{olley:1996}. Also, though not reported here, log(Material/Employees) interacted with $D^{0}$ and $1- D^{0}$ are additional explanatory variables when using the \citetalias{olley:1996} method, where $D^{0}$ is a dummy variable that takes value 1 if the firm reports zero R\&D expense.    Every specification includes Time and Industry Dummies.  }	\\															
				\legend  																	
		\end{tabular} }																
		\label{tabres3}																	
	\end{sidewaystable}

	\begin{table}[h!]	
		\caption{\textbf{Productivity Elasticity with respect to R\&D for  (a) firms in  High-Tech and Medium-High-Tech Manufacturing and (b) firms in the Remaining Sectors }}															
		\centering																	
		\begin{tabular}{l|r@{}l | r@{}l| r@{}l| r@{}l}																	
			\hline\hline															
			&	 \multicolumn{4}{c|}{Specification: Baseline} &	\multicolumn{4}{p{4.5cm}}{Specification: Baseline + External Knowledge}				\\\hline
			&\multicolumn{2}{c|}{Incumbent}& \multicolumn{2}{c|}{Entrant}& \multicolumn{2}{c|}{Incumbent}& \multicolumn{2}{c}{Entrant}\\\hline
			log(R\&D/Employees) $ \times D_{HM} $ & 0.061   & \onepc & 0.045   &  \fivepc &  0.094  & \onepc & 0.074 &\onepc \\
			& (0.002) &        & (0.021) &          & (0.006) &        & (0.005)  &       \\
			log(R\&D/Employees) $ \times (1-D_{HM}) $ & 0.037   & \onepc & 0.080   & \onepc   & 0.085   & \onepc & 0.091 & \onepc \\
			& (0.001) &        & (0.022) &          & (0.001) &        & (0.005) &      \\ \hline
			\multicolumn{9}{p{15.5cm}}{Test of Equality of Elasticity of Productivity with respect to R\&D for  (a) firms in  High-Tech and Medium-High-Tech Manufacturing and (b) firms in the Remaining Sectors}\\\hline
			Test Statistic: $ C $
			& 10.04   & \onepc & 1.51    &          &  68.4   & \onepc &  4.50 & \fivepc \\
			$ \pr ( \chi^{2}(1)>C) $ & (0.00)  &        & (0.22)  &          & (0.00)  &        &  (0.03) &       \\ \hline	\hline
			\multicolumn{9}{p{15.5cm}}{Note: $ D_{HM} $ is binary variable that takes value 1 if the firm is in the high-tech manufacturing or medium-high-tech manufacturing and 0 otherwise. For lack of space, we only display the productivity elasticities with respect to R\&D for (a) firms in the high-tech and medium-high-tech manufacturing and (b) firms in the remaining sectors. }\\	
			
			\legend  																	
		\end{tabular} 											
		\label{tabres4}																	
	\end{table}

	\begin{table}[h!]
		\caption{\textbf{Description of innovative activity: Product ($ \mathcal{P}^{D}_{jt} =1 $) and Process ($ \mathcal{P}^{C}_{jt} =1 $) Innovation}}
		\centering
		
		\begin{tabular}{l|c| c | c  }\hline \hline
			\multicolumn{4}{c}{Incumbents}\\\hline
			& $ \mathcal{P}^{C}_{jt} =1 $ & $ \mathcal{P}^{C}_{jt} =0 $
			& Total \\ \hline 
			$ \mathcal{P}^{D}_{jt} =1 $       &  1695   & 1650 & 3345 \\
			& (27.85) & (27.11) & (54.96) \\ \hline
			$ \mathcal{P}^{D}_{jt} =0 $      & 956 & 1785 & 2741  \\
			& (15.71) & (29.33) & (45.04) \\ \hline
			Total        & 2651 & 3435 & 6086 \\
			& (43.56) & (56.44) & (100)\\
			\hline \hline
		\end{tabular}
		\quad
		\begin{tabular}{l|c| c | c  }\hline \hline	
			\multicolumn{4}{c}{Entrants}\\\hline
			& $ \mathcal{P}^{C}_{jt} =1 $ & $ \mathcal{P}^{C}_{jt} =0 $
			& Total \\ \hline
			$ \mathcal{P}^{D}_{jt} =1 $ & 329     & 333  & 662        \\
			& (21.93) & (22.2) & (44.13)  \\ \hline
			$ \mathcal{P}^{D}_{jt} =0 $ & 260 &   578  & 838       \\
			& (17.33) & (38.53) & (55.87) \\ \hline
			Total & 589 &  911 & 1500        \\
			&(39.27) & (60.73) & (100)   \\\hline\hline
		\end{tabular}\label{table:ds4}
		\begin{flushleft}
			\hspace{1.15cm}Note: Cell percentage in parentheses.
		\end{flushleft}

	\end{table}

	\begin{table}[h!]	
		
		\caption{\textbf{Average Treatment Effect of Technological Innovation on Estimated Total Factor Productivity for Entrants and Incumbents}}															
		\centering																	
		\begin{tabular}{l|r@{}l|r@{}l|r@{}l|r@{}l }
			\hline \hline
			&\multicolumn{4}{c|}{Specification: Baseline }& \multicolumn{4}{c}{Specification: Baseline +  }\\
			&\multicolumn{4}{c|}{ }& \multicolumn{4}{c}{ External Knowledge }\\\hline
			
			& \multicolumn{2}{c|}{Incumbent} &  \multicolumn{2}{c|}{Entrant} & \multicolumn{2}{c|}{ Incumbent } & \multicolumn{2}{c}{Entrant}  \\ 
			\hline 
			ATE of  Product \textit{and/or}	Process & 0.129  & \fivepc & 0.267   &\onepc 	& 0.102  & \tenpc & 0.191   &\onepc 
			\\ 
			Innovation, $ I_{jt} $, on Productivity: $ \Delta $    & (0.051)&         & (0.072) &     & (0.054)&         & (0.079) &        \\
			\hline
			No. of matches for computing the ATE & 	\multicolumn{2}{c|}{1001} & \multicolumn{2}{c|}{378} &		\multicolumn{2}{c|}{1000} & \multicolumn{2}{c}{346}  \\ 
			\hline
			\hline
			ATE of \textit{only} Product	& 0.097  & \tenpc & 0.193   & 	& 0.116  & \fivepc & 0.135   & 
			\\ 
			Innovation on Productivity: $ \Delta_{10} $    & (0.057)&         & (0.123) &     & (0.058)&         & (0.196) &        \\
			\hline
			No. of matches for computing the ATE   & 	\multicolumn{2}{c|}{934} & \multicolumn{2}{c|}{123} &		\multicolumn{2}{c|}{910} & \multicolumn{2}{c}{114}  \\ 
			\hline \hline
			
			ATE of \textit{only} Process	& 0.048  &  & 0.227   & \fivepc
			& 0.066  &  & 0.223   & \fivepc 
			\\ 
			Innovation on Productivity: $ \Delta_{01} $    & (0.046)&         & (0.089) &     & (0.047)&         & (0.107) &        \\
			\hline
			No. of matches for computing the ATE & 	\multicolumn{2}{c|}{1286} & \multicolumn{2}{c|}{172} &		\multicolumn{2}{c|}{1250} & \multicolumn{2}{c}{158}  \\ 
			\hline
			\hline
			
			ATE of Product \textit{and} Process	& 0.134  & \onepc  & 0.305   & \onepc
			& 0.122  & \onepc  & 0.300   & \onepc 
			\\ 
			Innovation on Productivity: $ \Delta_{11} $    & (0.043)&         & (0.093) &     & (0.045)&         & (0.104) &        \\
			\hline
			No. of matches for computing the ATE  & 	\multicolumn{2}{c|}{1414} & \multicolumn{2}{c|}{198} &		\multicolumn{2}{c|}{1381} & \multicolumn{2}{c}{183}  \\

			\hline \hline
			Mean of Productivity &	\multicolumn{2}{c|}{7.496} & \multicolumn{2}{c|}{8.235} &	\multicolumn{2}{c|}{7.879} & \multicolumn{2}{c}{8.894} 
			\\ 
			
			\hline
			Standard Deviation of Productivity  &	\multicolumn{2}{c|}{0.742} & \multicolumn{2}{c|}{0.832}  &	\multicolumn{2}{c|}{0.735} & \multicolumn{2}{c}{0.815} 
			\\

			\hline \hline	
			\multicolumn{9}{p{16cm}}{Note: While estimating estimating $ \Delta $, we matched `exactly' on year dummies (2006 to 2018) and 2-digit industry dummies, and matched to `nearest neighbourhood' on the lags of the variables mentioned in the text. However, because there were much fewer firms who only innovated products, only innovated processes, or both, we could not match `exactly' on industry dummies while estimating $ \Delta_{10} $, $ \Delta_{01} $, and $ \Delta_{11} $. As a result of two different matching criteria, we find that for the incumbents, the number of matches are higher when estimating $ \Delta_{10} $, $ \Delta_{01} $, and $ \Delta_{11} $ even as the number of incumbents who only innovated products, or only innovated processes, or both were, by definition, smaller than the number of incumbents who innovated either products, or processes, or both.}	\\											
			\legend	
		\end{tabular} 														
		\label{tabres2}																	
	\end{table}																	
	
\end{document}